\documentclass[a4paper,fleqn,usenatbib]{mnras}

\usepackage[T1]{fontenc}
\usepackage{ae,aecompl}

\usepackage{graphicx}	% Including figure files
\usepackage{amsmath}	% Advanced maths commands
\usepackage{amssymb}	% Extra maths symbols
\usepackage{pdflscape}

\title[Gaia luminosities of B-type stars]{Recipes for bolometric corrections and \emph{Gaia} luminosities of B-type stars: Application to an asteroseismic sample}

\author[M. G. Pedersen et al.]{
May G. Pedersen,$^{1}$\thanks{E-mail: maygade.pedersen@kuleuven.be}
Ana Escorza,$^{1,2}$
P{\'e}ter I. P{\'a}pics$^{1}$
and Conny Aerts$^{1,3,4}$
\\
$^{1}$Institute of Astronomy, KU Leuven, Celestijnenlaan 200D, 3001 Leuven, Belgium\\
$^{2}$Institut d'Astronomie et d'Astrophysique, Universit{\'e} Libre de Bruxelles, Campus Plaine C.P. 226, Boulevard du Triomphe,\\ B-1050 Bruxelles, Belgium\\
$^{3}$Department of Astrophysics, IMAPP, Radboud University Nijmegen, P. O. Box 9010, 6500 GL Nijmegen, the Netherlands\\
$^{4}$Max Planck Institute for Astronomy, Koenigstuhl 17, 69117 Heidelberg, Germany
}

\date{Accepted 2020 April 30. Received 2020 April 30; in original form 2019 September 27}

\pubyear{2019}

\begin{document}
\label{firstpage}
\pagerange{\pageref{firstpage}--\pageref{lastpage}}
\maketitle

\begin{abstract}
We provide three statistical model prescriptions for the bolometric corrections appropriate for B-type stars as a function of: 1) $T_\text{eff}$, 2) $T_\text{eff}$, $\log g$, and 3) $T_\text{eff}$, $\log g$, [M/H]. These statistical models have been calculated for 27 different filters, including those of the \emph{Gaia} space mission, and were derived based on two different grids of bolometric corrections assuming LTE and LTE+NLTE, respectively. Previous such work has mainly been \textcolor{black}{limited to a single photometric passband without taking into account} NLTE effects on the bolometric corrections.
%mainly been done for cool stars with effective temperatures below 10000\,K, and/or has been limited to a single photometric passband without taking into account NLTE effects on the bolometric corrections.
Using these statistical models, we calculate the luminosities of 34 slowly pulsating B-type (SPB) stars with available spectroscopic parameters, to place them in the Hertzsprung-Russell diagram and compare their position to the theoretical SPB instability strip. We find that excluding NLTE effects has no significant impact on the derived luminosities for the temperature range \textcolor{black}{11500-21000\,K}\textcolor{black}{. W}e conclude that spectroscopic parameters are needed in order to achieve meaningful luminosities of B-type stars\textcolor{black}{. T}he three prescriptions for the bolometric corrections \textcolor{black}{are valid for any galactic} B-type star with \textcolor{black}{effective temperatures and surface gravities in the ranges 10000-30000\,K and 2.5-4.5\,dex, respectively}\textcolor{black}{, covering regimes below the Eddington limit.}
\end{abstract}


%%%%%%%%%%%%%%%%%%%%%%%%%%%%%%%%%%%%%%%%%%%%%%%%%%

%%%%%%%%%%%%%%%%% BODY OF PAPER %%%%%%%%%%%%%%%%%%

\section{Introduction}\label{sec:intro}

In order to derive stellar luminosities, one must know the total bolometric flux emitted by the star. This is a difficult quantity to measure and different approaches are usually taken to circumvent this problem. \textcolor{black}{Here, we focus on this problem for early-type stars  with effective temperatures $T_{\rm eff}\geq\,10^4$\,K. For such stars, the problem is often circumvented by working with} spectroscopic luminosities $L_\text{spec}$, which \textcolor{black}{are an approximation of the actual luminosities and} are calculated directly from the spectroscopic effective temperatures $T_\text{eff}$ and surface gravities $\log g$ \citep[$L_\text{spec}\equiv T_\text{eff}^4/\log g$,][]{Langer2014,Simon2017,Castro2018}. Bolometric luminosities can also be obtained by converting measured apparent magnitudes to absolute bolometric magnitudes using distance measurements and bolometric corrections, as previously done \textcolor{black}{ for stars with effective temperatures above $10000$\,K by, e.g., \citet{Humphreys1979,Underhill1980,Schonberner1984,Stahl1984,Singh1987,Massey1989a, Massey1989b,Parker1993,Hubrig2000,Hunter2007,Fossati2014,Camacho2016,Martins2019,Dufton2019,Dufton2020,Balona2019, Balona2020}.}

%\clearpage
Going from observed absolute magnitudes in a given passband $M_{S_\lambda}$ to absolute bolometric luminosities $L_\star$ requires a conversion from passband magnitudes to bolometric magnitudes through the use of bolometric corrections (BCs):

\begin{align}
-2.5 \log L_\star/L_\odot &= M_\text{bol} - M_{\text{bol},\odot}\nonumber\\
	&= M_{S_\lambda} + BC_{S_\lambda} - M_{\text{bol},\odot}.
	\label{Eq:Lum_BC_Mfilter}
\end{align}

\noindent Here $M_{\text{bol},\odot}$ and $L_\odot$ are the absolute bolometric magnitude and luminosity of the Sun, respectively. The $BC$ values are highly dependent on both a) the photometric passband used to carry out the observations, and b) the underlying stellar spectrum for which the correction has to be carried out. We represent the passband dependence through the subscript $S_\lambda$, which is the filter response as a function of wavelength $\lambda$ for a given photometric passband. 

The spectral dependence of the bolometric corrections represents itself as a dependence on the effective temperature, surface gravity, and metallicity [M/H] of the star, out of which $T_\text{eff}$ is the most important \textcolor{black}{parameter. For} stars with $T_\text{eff} \geq 15000$\,K, non-local thermodynamic equilibrium (NLTE) effects start to become important for the formation of spectral lines, and must be taken into account when deriving stellar abundances and spectroscopic parameters \citep[as done by, e.g.,][]{Morel2006,Hunter2007,Nieva2012}. The impact of including only local thermodynamic equilibrium (LTE) effects instead of NLTE at such high temperatures on the corresponding derived bolometric corrections, has so far not been investigated.

The derivation of bolometric corrections can be quite cumbersome if one has to rely on carrying out interpolations in existing grids \textcolor{black}{or tables} of bolometric corrections \textcolor{black}{\citep[e.g.][]{Kuiper1938,Morton1968,Flower1977,Hayes1978,Lanz1984,Malagnini1986,Chlebowski1991,Lanz2003,Lanz2007}}, or derive bolometric corrections from synthetic stellar spectra and pass\textcolor{black}{b}and transmission curves. Instead, prescription\textcolor{black}{s} for the bolometric corrections as a function of $T_\text{eff}$, $\log g$, and [M/H] provide a much faster way of deriving the bolometric corrections. \citet{Flower1996} provided three such prescriptions for calculating the bolometric corrections in the V passband from measured effective temperatures. The prescriptions are expressed as third to fifth order polynomials of $\log T_\text{eff}$ and are given for three different temperature ranges. They have been widely used in the stellar community \citep[e.g.][to name just a few recent examples]{Hanes2019,Walczak2019,Cunha2019,Sikora2019,Cokluk2019}. For the prescription valid for $\log T_\text{eff} \geq 3.90$, a fifth order polynomial was fit to get their statistical model (i.e. prescription) for $BC_\text{V}$. While this prescription is well-behaved for hot stars, it lacks the information of $\log g$ and [M/H] as well as errors on the coefficients of the polynomial fit. This leads to an underestimation of the errors on the final calculated luminosities.

\textcolor{black}{Aside from \citet{Flower1996}, several other attempts have been made at providing expressions for the bolometric correction in the V-band as a function of ($\log$)~$T_\text{eff}$. Some, like \citet{Flower1996}, provide their prescriptions for different temperature ranges like \citet{Massey1989a,Massey1989b} did, using first and second order polynomials of $\log T_\text{eff}$. Their prescriptions are based on the bolometric corrections tables by \citet{Flower1977}. Others use a single prescription over a wide range in temperature \citep[e.g.][third order polynomial for $BC_V$ as a function of effective temperature for stars earlier than G5]{Balona1994}, or provide linear prescriptions for stars with high effective temperatures as done by, e.g., \citet[][$T_\text{eff} > 30000$\,K]{Chlebowski1991}, \citet[][$\sim 28000-45000$\,K]{Martins2005}, and \citet[][for 15800-34000\,K]{Nieva2013}. Few attempts have been made at including also the surface gravity in the expressions for the bolometric correction (e.g. \citealt{Vacca1996} for stars with $T_\text{eff} \in [28200, 52500]$\,K), or for different photometric passbands (e.g. \citealt{Martins2006} for six different filters, UBVJHK, valid for stars with $T_\text{eff} > 25100$\,K). Except from \citet{Vacca1996}, none of the examples listed above provide errors on the regression coefficients but at most the standard deviations or root-mean-squared (rms) errors between the prescriptions and the bolometric correction values on which they are based (\citealt{Balona1994}: 0.047\,mag, \citealt{Nieva2013}: $0.01$\,mag, \citealt{Martins2006}: 0.05-0.10\,mag, \citealt{Martins2005}: 0.05\,mag.)
}

With the release of the \emph{Gaia} DR2 parallaxes and photometry \citep{GaiaCollaborationPrusti2016,GaiaCollaborationBrown2018,Evans2018,Lindegren2018b}, new extensive efforts are being made to derive accurate luminosities of stars across the Hertzsprung-Russell (HR) diagram. As was done by \citet{Flower1996}, \citet{Andrae2018} provided a prescription for doing so from the \emph{Gaia} photometry, by deriving two fourth order polynomial prescriptions for the bolometric corrections as a function of ($T_\text{eff} - T_{\text{eff},\odot}$) in the \emph{Gaia} G passband. They do so using a grid of bolometric corrections calculated from the MARCS synthetic spectra \citep{Gustafsson2008}, and provide errors on their estimated coefficients arising from the scatter in $\log g$. Both of the temperature ranges for which the two prescriptions are valid fall below $8000$\,K, because the MARCS models do not go to higher temperatures. Therefore none of their statistical models can be used for B-type stars, which are the focus of this work.

Granted the limited work that has previously been done in a) calculating prescriptions for the bolometric corrections as a function of $T_\text{eff}$, $\log g$, and [M/H] that are valid for B-type stars, b) doing so for a variety of different passbands, while c) accounting for NLTE effects, we aim \textcolor{black}{to provide} recipes for bolometric corrections to be used for calculating luminosities of B-type stars. These recipes are developed in order to place a newly selected sample of 34 slowly pulsating B-type (SPB) stars observed by the \emph{Kepler} space telescope \citep{Borucki2010} in the SPB instability strip in the HR diagram. This sample of B-type stars is unique in that they all have detected gravity mode period spacing series \citep[][]{Papics2014,Papics2015,Papics2017,Zhang2018,Szewczuk2018,Pedersen2020PhDT} and available spectroscopic parameters, making them prime targets for detailed asteroseismic modeling. Knowing the spectroscopic parameters as well as the luminosities, will serve as crucial constraints \textcolor{black}{for the asteroseismic modeling.} While the motivation behind this work is to obtain accurate luminosities of these 34~SPB stars, we stress this is just one out of many applications and that our prescriptions are valid for all stars with $T_\text{eff} \in [10000, 30000]$\,K in and near the main-sequence.

We calculate bolometric correction tables for both LTE and NLTE model atmospheres, and provide statistical model representations for the BC values in the temperature range 10000--30000\,K, using a multivariate linear regression scheme. The LTE and NLTE models are described in Sect.\,\ref{Sec:ModelAtm}, and the procedures for calculating the bolometric corrections are outlined in Sect.\,\ref{Sec:BolometricCorrections}. We construct two grids of bolometric corrections, one using solely the LTE models and another combining both LTE and NLTE into a LTE+NLTE grid of bolometric corrections, see Sect.\,\ref{Sec:BC_lte_vs_nlte}, in order to cover the required temperature range as well as compare the derived luminosities. \textcolor{black}{The impact of varying the microturbulence on the derived bolometric corrections is discussed in Sect.\,\ref{Sec:vturb}}. For both the LTE and LTE+NLTE grids, we derive three statistical model representations of the bolometric corrections (Sect.\,\ref{Sec:BC_stat_model}). The first statistical model includes only the effective temperature, while the second and third also includes the $\log g$ and [M/H]. Section\,\ref{Sec:Lum_derivarion} outlines how the luminosities are derived from an averaged, extinction corrected, apparent bolometric magnitude based on these prescriptions of the bolometric corrections. We investigate in Sect.\,\ref{Sec:Lum_comparison} how the luminosities change depending on a) if the LTE or LTE+NLTE grids are used in the derivation of the statistical models, b) the \emph{Gaia} distances from \citet{Bailer-Jones2018} or \citet{Anders2019} are applied, \textcolor{black}{and c)} the spectroscopic parameters are replaced by the \emph{Gaia} $T_\text{eff}$, $\log g$, [M/H] parameters from \citet{Anders2019} or by those from the \emph{Kepler Input Catalog} \citep[KIC,][]{KIC2009}. Furthermore, we check how our derived luminosities compare to the ones for which the \citet{Flower1996} prescription for the $\text{BC}_V$ has been used in the derivation of the bolometric corrections, and if an interpolation on the LTE+NLTE grids have been made to achieve the bolometric corrections. Finally, we place the 34~SPB stars in the HR diagram in Sect.\,\ref{Sec:HRD} to compare with the theoretical SPB instability strip from \citet{Moravveji2016}, and present our conclusions in Sect.\,\ref{Sec:Conclusions}.

%\clearpage
\section{Choice of model atmospheres}\label{Sec:ModelAtm}

\begin{figure}%[h]
  \centering
  \includegraphics[width=\linewidth]{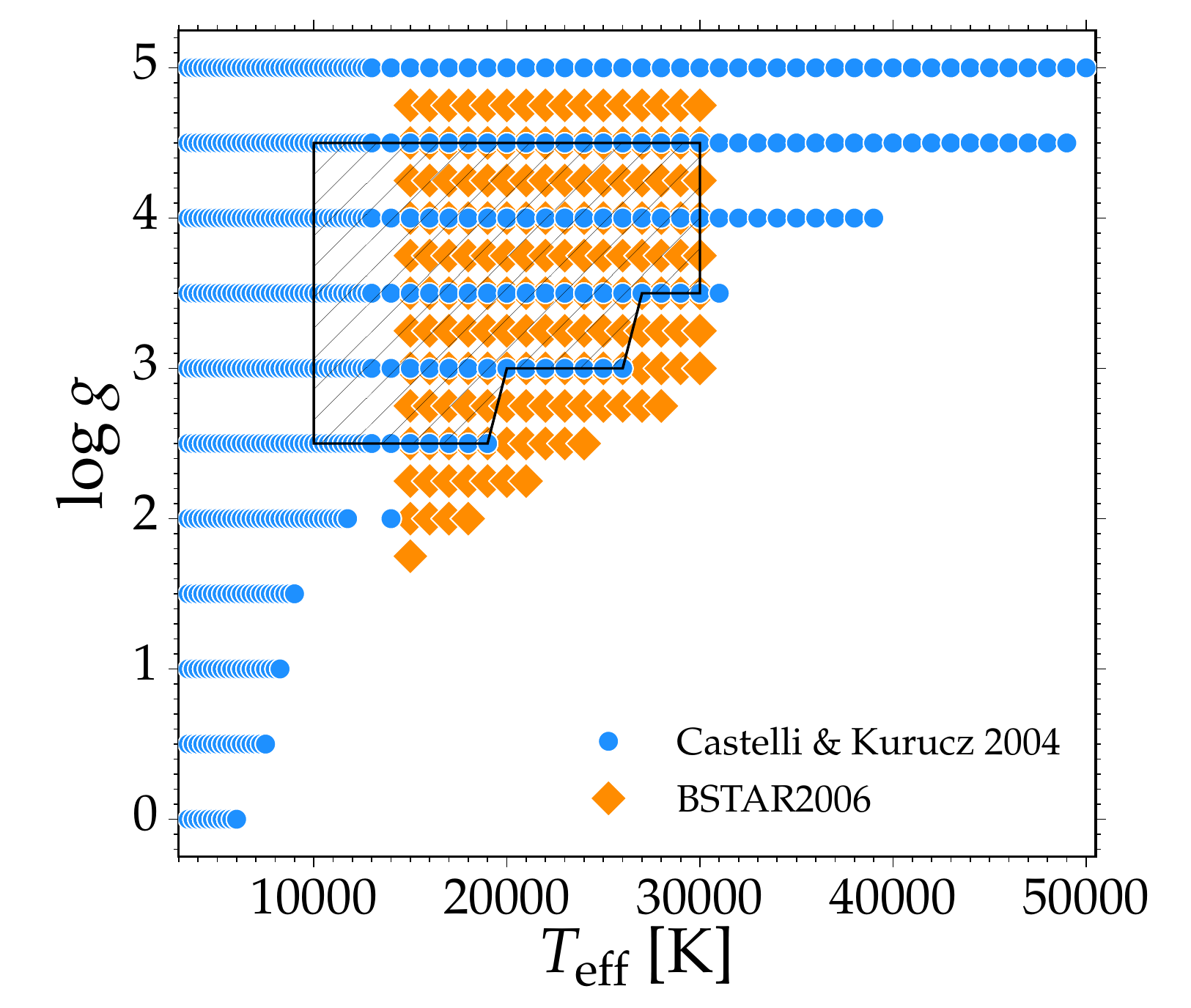}
\caption{Parameter space in $T_\text{eff}$ and $\log g$ covered by the two grids of model atmospheres considered in this work. The hatched region indicates the parameter range used to derive a statistical model for the bolometric corrections.}
	\label{fig:GridParameters}
\end{figure}

In order to investigate the impact of taking NLTE effects into account on the derived bolometric corrections, we consider two grids of spectral energy distributions (SEDs): the ATLAS9 \citet[][LTE]{Castelli2003,Castelli2004} and TLUSTY BSTAR2006 \citep[][NLTE]{Lanz2007} models. Both grids are described in detail below, and their coverage in $T_\text{eff}$ and $\log g$ are shown in Fig.\,\ref{fig:GridParameters}. \textcolor{black}{All these considered SED models have scaled solar abundances, thus we warn against extrapolating to stars where this is a poor approximation.}

%\clearpage
\subsection{ATLAS9 Kurucz 2004}

The grid of ATLAS9 stellar model atmospheres by \citet{Castelli2003,Castelli2004} is calculated assuming LTE. It covers a range in effective temperatures from 3000-50000\,K and surface gravity of $\log g \in [0.0, 5.0]$ in steps of 0.5\,dex, as shown in Fig.\,\ref{fig:GridParameters} by the blue, filled circles. At each of these grid points, models are available for eight different metallicities, with [M/H]\,$\in [-2.5, 0.0]$ in steps of 0.5\,dex and two higher values at +0.2 and \textcolor{black}{+}0.5. The microturbulent velocity has been fixed to $V_t = $2\,km\,s$^{-1}$, and the solar abundances of \citet{Grevesse1998} are assumed in the calculations. The SEDs cover a wavelength range of \textcolor{black}{$90-1600000$}\,\AA\ and the fluxes are given in units of $F_\lambda$, i.e. ergs\,cm$^{-2}$\,s$^{-1}$\,\AA$^{-1}$. For more details on the computation of the models we refer the reader to \citet{Castelli2003,Castelli2004}. \textcolor{black}{The ATLAS9 SED models can be downloaded from here: \url{https://archive.stsci.edu/hlsps/reference-atlases/cdbs/grid/ck04models/}.}

%\clearpage
\subsection{TLUSTY BSTAR2006}

BSTAR2006 is a grid of NLTE model atmospheres calculated using the TLUSTY code \citep{Hubeny1995,Lanz2007}. The grid covers a range in $T_\text{eff}$ of 15000-30000\,K in steps of 1000\,K, and 16 surface gravities of $\log g \in [1.75, 4.75]$ in steps of 0.25\,dex, as shown in Fig.\,\ref{fig:GridParameters} by the orange diamonds. Models are available for five metallicities at each grid point: $Z/Z_\odot$ = 2, 1, 1/2, 1/5, and 1/10, i.e. [M/H] = 0.3, 0.0, -0.3, -0.7, and -1.0. Like for the \citet{Castelli2003,Castelli2004} models, the solar abundances of \citet{Grevesse1998} and $V_t = $2\,km\,s$^{-1}$ are assumed. The BSTAR2006 grid also provides models for $V_t = $10\,km\,s$^{-1}$ \textcolor{black}{when $\log g \leq 3.0$. The impact of increasing the microturbulence from 2\,km\,s$^{-1}$ to 10\,km\,s$^{-1}$ on the bolometric corrections will be investigated separately. We include only the SED models with $V_t = $2\,km\,s$^{-1}$ in the LTE and NLTE comparisons, the derivation of bolometric correction prescriptions, and the final derived stellar luminosities.
The SED models in the BSTAR2006 grid} cover a wavelength range of $55\textcolor{black}{-}3000000$\,\AA\ and are provided in units of $F_\nu$ (i.e. $\text{erg} \ \text{cm}^{-2} \ \text{s}^{-1} \ \text{Hz}^{-1}$) as a function of frequency $\nu$ in s$^{-1}$. Before use, $F_\nu$ is converted to $F_\lambda$ and $\nu$ to $\lambda$ \textcolor{black}{such that} the BSTAR2006 SEDs have the same units as the ones of \citet{Castelli2003,Castelli2004}. The BSTAR2006 SED models used in this work have been downloaded from \url{http://tlusty.oca.eu/Tlusty2002/tlusty-frames-BS06.html}.

%\clearpage
\section{From model atmospheres to bolometric corrections}\label{Sec:BolometricCorrections}

\textcolor{black}{The} bolometric correction $BC_{S_\lambda}$ needed to convert passband magnitudes into their corresponding bolometric counterpart \textcolor{black}{depends on} the underlying stellar spectrum as well as \textcolor{black}{on} the photometric passband $S_\lambda$. Therefore, when deriving $BC_{S_\lambda}$ one generally relies on grids of synthetic stellar spectra for varying $T_\text{eff}$, $\log g$ and [M/H] and derives a grid of $BC_{S_\lambda}$ values for different combinations of these parameter \citep[see also previous work by, e.g., ][]{Bessell1998,Girardi2002,Bell2014,Casagrande2018a}. Due to different wavelength coverage and transmission efficiency for different filters, see Fig.\,A1 \textcolor{black}{in Appendix~A} \textcolor{black}{available online}, each considered photometric passband requires a $BC_{S_\lambda}$ grid of its own. \textcolor{black}{Table\,A1 in the appendix} provides an overview of the filter properties of the photometric passbands for which photometric data is available for the 34 SPB stars considered in this work.

%\clearpage
\subsection{Equation for bolometric correction and photometric zero points}\label{Sec:BCeqZPeq}
 
The \textcolor{black}{applied} equation for bolometric correction \textcolor{black}{in a given photometric passband with response function $S_X$} is based on previous work by \citet{Bessell1998,Girardi2002,Bessell2012}:

\begin{align}
BC_{S_X} = \  &2.5 \log\left[ \frac{\int F_X S_X X dX}{\int S_X X dX}\right] - 2.5\log\left[T_\text{eff}^4\right]\nonumber\\ 
&+ \text{const.} + zp.
	\label{Eq:BC}
\end{align}

\noindent \textcolor{black}{Here} $\text{const.} = M_{\text{bol},\odot} - 2.5 \log\left[4 \pi \sigma \left(10 \text{pc} \right)^2/L_\odot\right] = -0.8814$, using $M_{\text{bol},\odot} = 4.74$ and $L_\odot = 3.828 \times 10^{33} \ \text{erg/s}$\footnote{$M_{\text{bol},\odot}$ and $L_\odot$ are taken from the IAU resulution 2015 B2, \url{https://www.iau.org/static/resolutions/IAU2015_English.pdf}}, \textcolor{black}{and $zp$ is the zero-point of the given magnitude system}. This equation \textcolor{black}{is listed in its} most general form, where $X$ is either $\lambda$ or $\nu$ (i.e. wavelength [\AA] or frequency [Hz]) \textcolor{black}{for the} VEGAmag system ($X = \lambda$) and the ABmag system ($X = \nu$). \textcolor{black}{$F_X$ is the flux density.}

In the ABmag system $zp = 48.60$ per definition. In comparison, the VEGAmag system is defined such that all colours are zero and the apparent magnitude of Vega is $V = 0.03$, i.e. $m_{S_\lambda}^\text{Vega} = 0.03$. To make sure that this is always fullfilled, \textcolor{black}{$zp$ becomes} passband dependent. Here we calculate $zp$ for the VEGAmag system for each passband using

\begin{equation}
zp = -2.5 \log \left[ \frac{\int f^\text{Vega}_\lambda S_\lambda \lambda d\lambda}{\int S_\lambda \lambda d\lambda}\right] - m_{S_\lambda}^\text{Vega}.
	\label{AEq:ZP}
\end{equation}

\noindent In the case of the Stromgren passbands, $m_{S_\lambda}^\text{Vega} = 1.432, 0.179, 0.018, 0.014$ are used for the Stromgren $UVBY$ filters, respectively \citep{MaizApellaniz2007}.
For the spectrum of Vega measured at the Earth, $f^\text{Vega}_\lambda$, we use \emph{alpha\_lyr\_stis\_008.fits} which has been normalised to $3.44 \times 10^{-9} \text{erg} \ \text{cm}^{-2} \ \text{s}^{-1}$\AA$^{-1}$ at $5556$\,\AA \ and is available at the CALSPEC database\footnote{http://www.stsci.edu/hst/observatory/crds/calspec.html} \citep{Bohlin2004,Bohlin2007,Bohlin2014}. The calculated  $zp$-values are listed in Table\,A1.

The equations given above are listed assuming that the detector type is a photon counter. As discussed by \citet{Girardi2002}, for an energy counter detector all $X dX$ terms in \textcolor{black}{Eq.\,(\ref{Eq:BC})} simplify to $dX$, and $\lambda d\lambda$ is replaced by $d\lambda$ in Eq.\,(\ref{AEq:ZP}).

\subsection{Absolute magnitudes of the Sun}\label{Sec:AbsSun}

\begin{figure*}%[h]
  \centering
  \includegraphics[width=\linewidth]{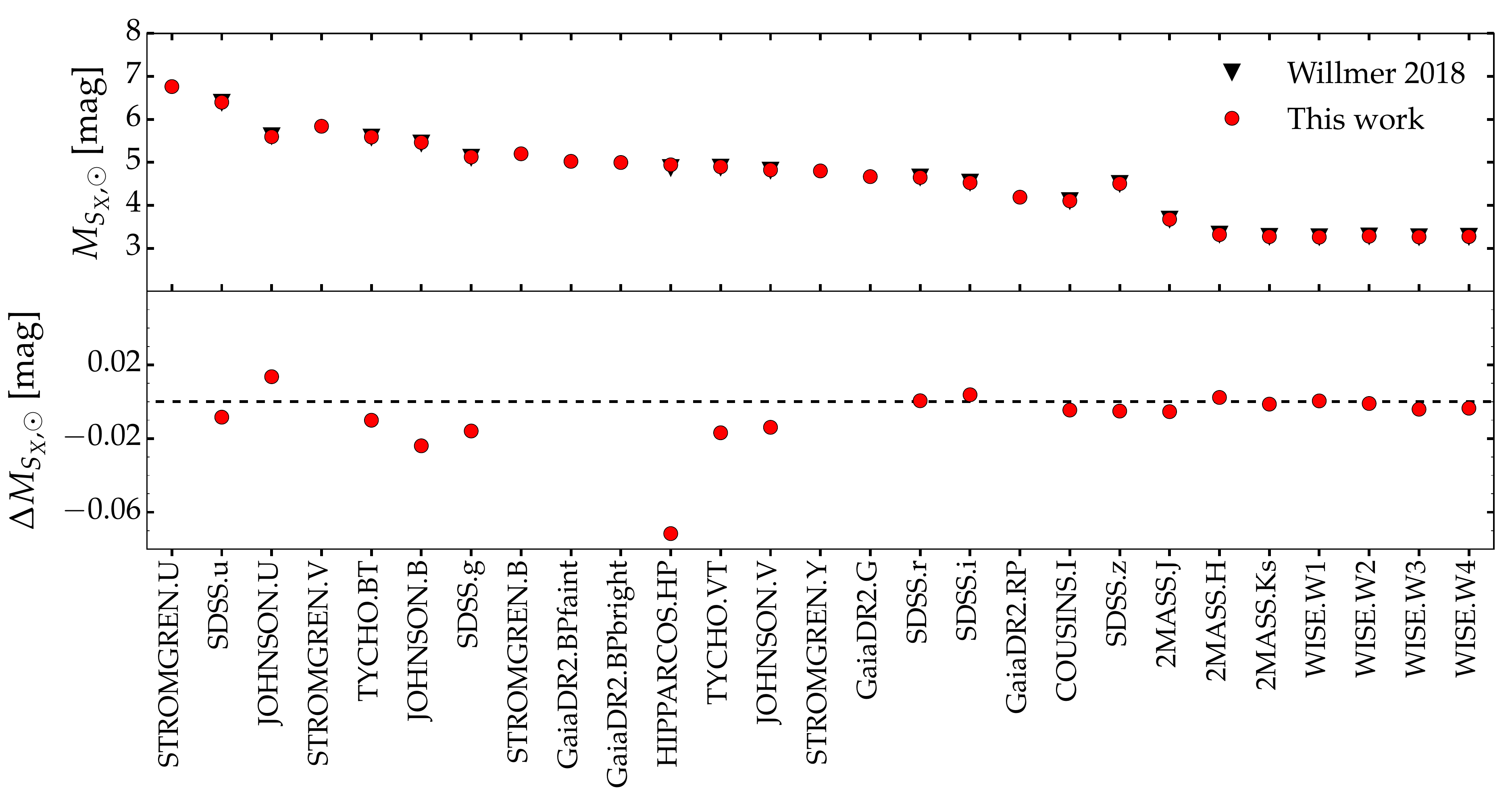}
\caption{Absolute magnitudes of the Sun (\emph{top}) calculated in different filters specified on the x-axis and ordered according to their effective wavelength. The results by \citet{Willmer2018} are shown in black, inverted triangles, while the red circles denote the values calculated in this work. The bottom panel shows the differences where available. Except for the SDSS filters, all magnitudes are given in VEGAmag.}
	\label{fig:WillmerVsPedersen}
\end{figure*}

As discussed extensively by \citet{Torres2010}, when using bolometric corrections calculated from synthetic spectra it is important to make sure that these are consistent with $M_{\text{bol},\odot}$. \textcolor{black}{We} choose to calculate the expected absolute magnitudes of the Sun using the solar composite spectrum observed and calculated by \citet{Haberreiter2017}. This has previously been done by \citet{Willmer2018} for a large sample of photometric passbands. However, because the absolute magnitude of the Sun in the \textit{Gaia} and Stromgren passbands were not included by \citet{Willmer2018} and in order to be consistent, this procedure is repeated here. 

The absolute magnitudes of the Sun, $M_{S_X,\odot}$, in the different photometric passbands used in this work are calculated using the equation

\begin{equation}
M_{S_X, \odot} = -2.5 \log \left[ \frac{\int f^\text{Sun}_X S_X X dX}{\int S_X X dX}\right] - zp -  5 \log\left[1 \ \text{AU}\right] + 5,
	\label{Eq:MsunFilt}
\end{equation}

\noindent with $1 \ \text{AU} = 4.8481\times 10^{-6} \ \text{pc}$. \textcolor{black}{From} these absolute magnitudes and $M_{\text{bol},\odot}$ we derive the required bolometric correction to reproduce $M_{S_X, \odot}$. The calculated $BC_{S_X}$ grids are subsequently shifted such that $BC_{S_X, \odot} = M_{\text{bol},\odot} - M_{S_X, \odot}$ is obtained at $T_{\text{eff}, \odot} = 5772 \ \text{K}$, $\log g_\odot = 4.438$ and [M/H]~=~0.

Figure\,\ref{fig:WillmerVsPedersen} shows the difference between our calculated absolute solar magnitudes (red circles) and those by \citet{Willmer2018} (black, inverse triangles). \textcolor{black}{A} difference between the values derived in this work and those of \citet{Willmer2018} is that \citet{Willmer2018} assumes photon counting detectors in their calculations. In comparison, we adjust Eq.\,(\ref{Eq:MsunFilt}) depending on if the passband transmission curves are given for a photon or energy counter detector, the latter being the case for the majority of the filters.

\subsection{LTE vs NLTE bolometric corrections}\label{Sec:BC_lte_vs_nlte}

\begin{figure*}
  \centering
  \includegraphics[width=\linewidth]{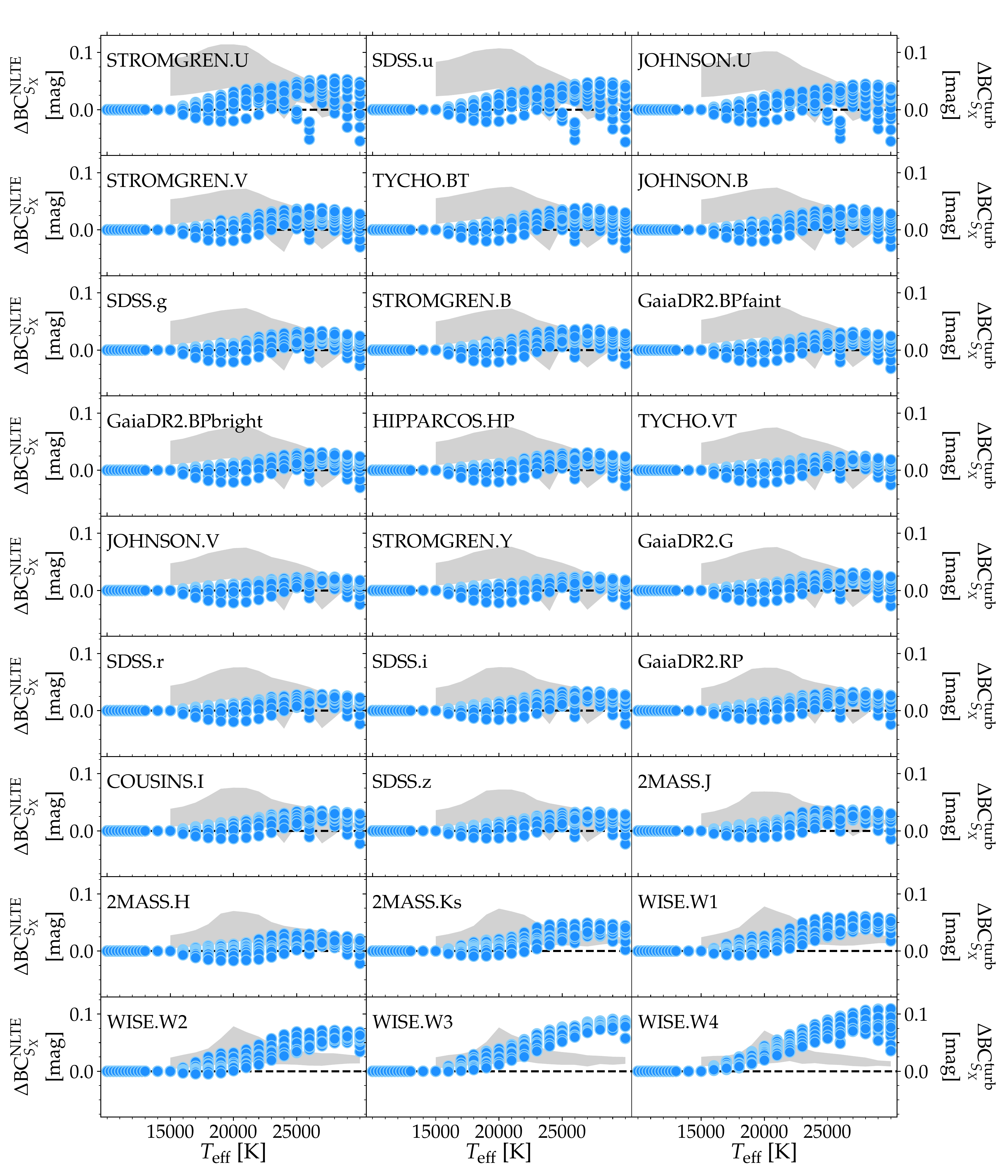}
\caption{Differences between the bolometric corrections derived for the grid of LTE models and the combined LTE+NLTE bolometric corrections grid in the different passbands. $\Delta$BC$^\text{NLTE}_{S_X}$ = BC$^\text{LTE+NLTE}_{S_X}$ - BC$^\text{LTE}_{S_X}$. The \textcolor{black}{black} dashed line shows the position of $\Delta$BC$^\text{NLTE}_{S_X}$ = 0. \textcolor{black}{For comparison we likewise plot as grey shaded regions the differences $\Delta$BC$^\text{\,turb}_{S_X}$ resulting from using a microturbulence parameter of 10\,km\,s$^{-1}$ instead of 2\,km\,s$^{-1}$ for $\log g \in [2.5, 3.0]$ in the BSTAR2006 NLTE grid. $\Delta$BC$^\text{\,turb}_{S_X}$ = BC$^{\,10\,\text{km\,s}^{-1}}_{S_X}$ - BC$^{\,2\,\text{km\,s}^{-1}}_{S_X}$.}}
	\label{fig:lte_vs_nlte}
\end{figure*}

In order to investigate the effects of using SEDs with NLTE effects taken into account on the derived luminosities, we calculate two grids of bolometric corrections. The first grid uses the SED models of \citet{Castelli2003,Castelli2004} with $\log g \geq 2.5$\textcolor{black}{, i.e. all models indicated by blue circles in Fig.\,\ref{fig:GridParameters} with $\log g \geq 2.5$,} and will be referred to as the LTE grid henceforth. Because the BSTAR2006 grid only starts at 15000\,K but the relevant temperature range for SPB stars starts at 10000\,K, the second grid is a combination of the bolometric corrections derived from the \citet{Castelli2003,Castelli2004} and the BSTAR2006 models. We refer to this grid as the LTE+NLTE grid.

\textcolor{black}{For} bolometric corrections for temperatures below 15000\,K, the values derived for the LTE grid are used. These values have been adjusted such that the absolute magnitude of the Sun can be reproduced, see Sect.\,\ref{Sec:AbsSun}. At higher temperatures, the bolometric corrections are calculated for all grid points in the BSTAR2006 grid. 
\textcolor{black}{As the sampling in metallicity and surface gravity is different for the two grids (cf. Fig.\,\ref{fig:GridParameters}), this may lead to an artificial weighting in the calculation of the prescriptions for the bolometric corrections. To circumvent this, we carry out a cubic interpolation in order to map the bolometric corrections of the BSTAR2006 grid onto the covered $\log g$ and [M/H] parameters in the \citet{Castelli2003,Castelli2004} LTE grid within the temperature range 15000-30000\,K,}
%Because the covered values in $\log g$ and [M/H] are different for the two grids of SED models, a cubic interpolation is carried out for the \citet{Castelli2003,Castelli2004} parameters in the 15000-30000\,K temperature range onto the BSTAR2006 bolometric correction grid, 
where the $\log g$ and [M/H] overlap in parameter space between the two \textcolor{black}{grids}. Because the BSTAR2006 grid covers a smaller range in $\log g$ and [M/H] values, the final parameter range of the LTE+NLTE grid becomes $T_\text{eff} \in [3500, 30000]$\,K, $\log g \in [2.5, 4.5]$\footnote{\textcolor{black}{The exact allowed combination of $T_\text{eff}$ and $\log g$ is set by the Eddington limit, and for $20000\,\text{K} \leq T_\text{eff} \leq 30000$\,K no bolometric correction estimates exists for $\log g < 3.0$ in either the LTE or the LTE+NLTE grid.}}, and [M/H]\,$\in [-1, 0.2]$. For the sake of comparison\textcolor{black}{,} we restrict the LTE grid to this parameter range\textcolor{black}{. The} interpolated bolometric corrections in the LTE+NLTE grid at $T_\text{eff} \geq 15000$\,K are shifted such that they have the same value at 15000\,K as in the LTE grid. \textcolor{black}{This final shift is done in order to account for the fact that the bolometric corrections in the LTE+NLTE grid below 15000\,K were adjusted in order to reproduce the absolute magnitude of the Sun, while this had not been accounted for in the BSTAR2006 grid.}

The differences between the bolometric corrections derived using the \citet{Castelli2003,Castelli2004} LTE and the combined LTE+NLTE grid are shown in Fig.\,\ref{fig:lte_vs_nlte} for each passband\textcolor{black}{:}

\begin{equation}
\Delta\text{BC}_{S_X}^\text{NLTE} = \text{BC}^\text{LTE+NLTE}_{S_X} - \text{BC}^\text{LTE}_{S_X}, 
\end{equation}

\noindent and are shown for all grid points as a function of the effective temperature. Until the stitching point at $T_\text{eff} = 15000$\,K the differences are equal to zero by construction. At higher temperatures, $\Delta\text{BC}_{S_X}^\text{NLTE}$ is largely passband dependent and becomes as large as $\sim 0.1$\,dex at $T_\text{eff} = 30000$\,K for the WISE passbands. For temperatures below $T_\text{eff} = 25000$\,K, $\Delta\text{BC}_{S_X}^\text{NLTE}$ is less than $\sim \pm 0.05$\,dex for all but the WISE $W2$, $W3$, and $W4$ passbands and comparable in scale to the errors on the observed magnitudes of the stars. We discuss the impact on using the LTE and LTE+NLTE bolometric corrections on the calculated luminosities in Sect.\,\ref{Sec:Lum_comparison}.

\subsection{\textcolor{black}{Impact of microturbulence}}\label{Sec:vturb}

\textcolor{black}{Microturbulence is known to cause a broadening of spectral lines as well affect the effective temperature of the star and thereby its flux \citep[see e.g.,][for a detailed discussion on the relation between $T_\text{eff}$ and $V_t$]{Tkachenko2020}. It has been shown to vary as a function of the surface gravity for OB-type stars, increasing in value for both giants and supergiants (cf. e.g. Fig.~7 in \citealt{Cantiello2009} based on data from the ESO VLT-FLAMES Survey of Massive Stars). With this in mind, we investigate how such an increase in the microturbulence impacts the resulting bolometric corrections. Here we are limited by the parameter ranges covered in both the ATLAS9 and BSTAR2006 SED models, where the BSTAR2006 grid is the only one of the two which has models computed for different $V_t$ values. Furthermore, these SED models are only available for two different $V_t$ values (2 and 10\,km\,s$^{-1}$) and only when $\log g \leq 3.0$.}

\textcolor{black}{In the comparison of the bolometric correction predictions from $V_t=2$ and 10\,km\,s$^{-1}$ we include all BSTAR2006 SED models with $T_\text{eff} \in [15000, 30000]$ which have $\log g$ values between 2.5 and 3.0\,dex (i.e. models at the position of the orange diamonds between $\log g = 2.5$ and 3.0\,dex in Fig.~\ref{fig:GridParameters}) as well as the full range in metallicity. The resulting range in deviations $\Delta$BC$^\text{\,turb}_{S_X}$ = BC$^{\,10\,\text{km\,s}^{-1}}_{S_X}$ - BC$^{\,2\,\text{km\,s}^{-1}}_{S_X}$ across all included $\log g$ and [M/H] values are shown as a function of $T_\text{eff}$ by the grey shaded regions in Fig.~\ref{fig:lte_vs_nlte} for each passband\footnote{\textcolor{black}{BC$^{\,10\,\text{km\,s}^{-1}}_{S_X}$ and BC$^{\,2\,\text{km\,s}^{-1}}_{S_X}$ corresponds to the bolometric correction predictions using SED models with $V_t=2$ and 10\,km\,s$^{-1}$, respectively.}}. The $V_t= 10$\,km\,s$^{-1}$ models generally predict higher BC$_{S_X}$ values particularly at lower effective temperatures. The deviations are larger in the U passbands and decreases as we move towards filters at longer effective wavelengths. Below $\sim 20000$\,K, the $\Delta$BC$^\text{\,turb}_{S_X}$ values are larger than the differences resulting from the LTE vs LTE+NLTE comparisons discussed above, and move towards similar or smaller values at higher effective temperatures.}

%\clearpage
\subsection{Statistical model representation}\label{Sec:BC_stat_model}

The use of statistical model representations to determine bolometric corrections comes at several advantages. They provide an easy and quick way of calculating bolometric corrections based on other observed parameters, without having to rely on the availability of grids from which the corrections can be extracted through interpolations. Furthermore, the errors on the bolometric corrections can directly be obtained through normal error propagation of both the errors on the parameters and on the estimated coefficients of the statistical model. 

Because the vast majority of the passbands listed in Table\,A1 lack a prescription for the bolometric corrections and/or errors on the estimated coefficients, we calculate statistical model representations for $BC_{S_\lambda}$ using our previously calculated LTE and LTE+NLTE grids. We consider the three variables: $x_1 = \log T_\text{eff}/T_\text{eff,0}$, $x_2 = \log g$, and $x_3 = \text{[M/H]}$, where $T_\text{eff,0} = 10000$\,K is the lower \textcolor{black}{end for} which the statistical model will be computed. \textcolor{black}{We} construct three different statistical model representations of $BC_{S_\lambda}$, from which the bolometric corrections can be derived depending on what information is available for a given \textcolor{black}{star:}

\begin{align}
\text{Model 1:} \ BC_{S_\lambda} = \beta_0 &+ \beta_1 x_1 + \beta_2 x_1^2 + \beta_3 x_1^3
	\label{Eq:Model1}
\end{align}

\begin{align}
\text{Model 2:} \ BC_{S_\lambda} = \beta_0 &+ \beta_1 x_1 + \beta_2 x_1^2 + \beta_3 x_1^3\nonumber\\
&+ \beta_4 x_2  + \beta_5 x_2^2
	\label{Eq:Model2}
\end{align}

\begin{align}
\text{Model 3:} \ BC_{S_\lambda} = \beta_0 &+ \beta_1 x_1 + \beta_2 x_1^2 + \beta_3 x_1^3\nonumber\\
&+ \beta_4 x_2  + \beta_5 x_2^2\nonumber\\
&+ \beta_{6} x_3 + \beta_{7} x_3^2
	\label{Eq:Model3}
\end{align}

\noindent In order to determine the $\beta$'s in these three linear models, we carry out a multivariate linear regression \textcolor{black}{based on all bolometric correction estimates in the LTE and LTE+NLTE grids} within the temperature range $T_\text{eff} \in [10000, 30000]$\,K using the public statistics software package \textsc{R} \citep{Rcode}. We do so by first requiring that the first two orders of $x_1$, $x_2$, and $x_3$ are included in the regression\footnote{For the STROMGREN.U, SDSS.u, and JOHNSON.U passbands we also require that the third order term on $x_1$ is included in the multivariate linear regression.}. Then we check if adding higher order terms improves the fit by comparing the resulting Bayesian Information Criterion \citep[BIC;][]{schwarz1978} values of the models, and keep the one that achieves the smallest BIC value. This is done independently for each passband, allowing the included terms to be different depending on the passband, model, and bolometric correction grid used in the multivariate linear regression. The parameter range included in the regression is indicated by the black, hatched region in Fig.\,\ref{fig:GridParameters}\textcolor{black}{, indicating the region in $T_\text{eff}$ and $\log g$ for which our statistical models are valid. We summarize these ranges in Table~\ref{Tab:validity_range} and emphasize that the prescriptions provided in this work should not be used outside of these ranges. The corresponding validity range in metallicity is [M/H]$\, \in [-1.0, 0.2]$.} 

\begin{table}
	\caption{\textcolor{black}{Validity ranges in $T_\text{eff}$ and $\log g$ of the derived statistical models.}}
	\centering
	\label{Tab:validity_range}
	\begin{tabular}{cc}
	\hline\\[-1.5ex]
		\text{$T_\text{eff}$ [K]}	&	\text{$\log g$}\\
	\hline\\[-1.5ex]
		$[10000, 20000[$		&	$[2.5,4.5]$\\[0.5ex]
		$[20000,27000[$		&	$[3.0,4.5]$\\[0.5ex]
		$[27000,30000]$		&	$[3.5,4.5]$\\[0.5ex]
	\hline
	\end{tabular}
	\vspace{1ex}\\
	
	\raggedright\small \textcolor{black}{\textbf{Notes:} For all three listed combinations of $T_\text{eff}$ and $\log g$, the corresponding validity range in metallicity is [M/H]$\, \in [-1.0, 0.2]$.}
\end{table}

In this forward modelling approach, we have checked all terms of $x_1$, $x_2$, and $x_3$ up to the sixth order and find that for all three models, the highest number of relevant orders according to the BIC values is three, two, and two for $x_1$, $x_2$, and $x_3$, respectively. Hence, the higher order terms have not been included in Eq.(\ref{Eq:Model1})-(\ref{Eq:Model3}). For the \emph{Gaia} filters, the final determined model 3 representations of the LTE+NLTE grid are \textcolor{black}{provided} in Table\,\ref{Tab:beta_gaia}. The final derived $\beta$ coefficients and their errors are \textcolor{black}{tabulated in Appendix}\,D \textcolor{black}{available online} for all \textcolor{black}{27} passbands, the three statistical models and both the LTE and LTE+NLTE grid. \textcolor{black}{By comparing the final BIC values of Eq.\,(\ref{Eq:Model1})-(\ref{Eq:Model3}) we find that the statistical model that includes all three parameters $T_\text{eff}$, $\log g$, and [M/H] does a better job at representing the grids of bolometric corrections than the models including only $T_\text{eff}$ and/or $\log g$ independently of the photometric passband, even when punishing for including a higher number of fitting parameters. Figure D1 in Appendix~D illustrates the statistical model 3 representation of the LTE+NLTE $BC_{S_\lambda}$ grid for every passband.}

\begin{table*}
	\caption{Derived coefficients for the statistical model\,3 for the GaiaDR2.G, GaiaDR2.BPfaint, GaiaDR2.BPbright, and GaiaDR2.RP passbands using the LTE+NLTE grid.}
	\centering
	\label{Tab:beta_gaia}
	\begin{tabular}{ccccc}
	\hline\\[-1.5ex]
		\text{Coefficient}	&	\text{GaiaDR2.$G$}	&	\text{GaiaDR2.$BPfaint$}	&	\text{GaiaDR2.$BPbright$}	&\text{GaiaDR2.$RP$}\\
	\hline\\[-1.5ex]
		$\beta_{0}$	&	-0.2986$\pm$0.0258	&	-0.2820$\pm$0.0254	&	-0.3021$\pm$0.0255	&	0.0293$\pm$0.0058\\[0.5ex]
		$\beta_{1}$	&	-5.4963$\pm$0.0231	&	-5.0094$\pm$0.0227	&	-5.1276$\pm$0.0227	&	-6.1350$\pm$0.0240\\[0.5ex]
		$\beta_{2}$	&	-0.0257$\pm$0.0500	&	-0.2362$\pm$0.0492	&	-0.1952$\pm$0.0493	&	0.2026$\pm$0.0520\\[0.5ex]
		$\beta_{3}$	&	0.0000$\pm$0.0000	&	0.0000$\pm$0.0000	&	0.0000$\pm$0.0000	&	0.0000$\pm$0.0000\\[0.5ex]
		$\beta_{4}$	&	0.0692$\pm$0.0149	&	0.0788$\pm$0.0147	&	0.0836$\pm$0.0147	&	0.0264$\pm$0.0155\\[0.5ex]
		$\beta_{5}$	&	-0.0127$\pm$0.0021	&	-0.0147$\pm$0.0021	&	-0.0149$\pm$0.0021	&	-0.0074$\pm$0.0022\\[0.5ex]
		$\beta_{6}$	&	0.0994$\pm$0.0050	&	0.1088$\pm$0.0049	&	0.1070$\pm$0.0049	&	0.0837$\pm$0.0052\\[0.5ex]
		$\beta_{7}$	&	0.0293$\pm$0.0058	&	0.0330$\pm$0.0057	&	0.0325$\pm$0.0057	&	0.0219$\pm$0.0060\\[0.5ex]
	\hline
	\end{tabular}
\end{table*}

\begin{figure}
\centering
  \includegraphics[width=0.95\linewidth]{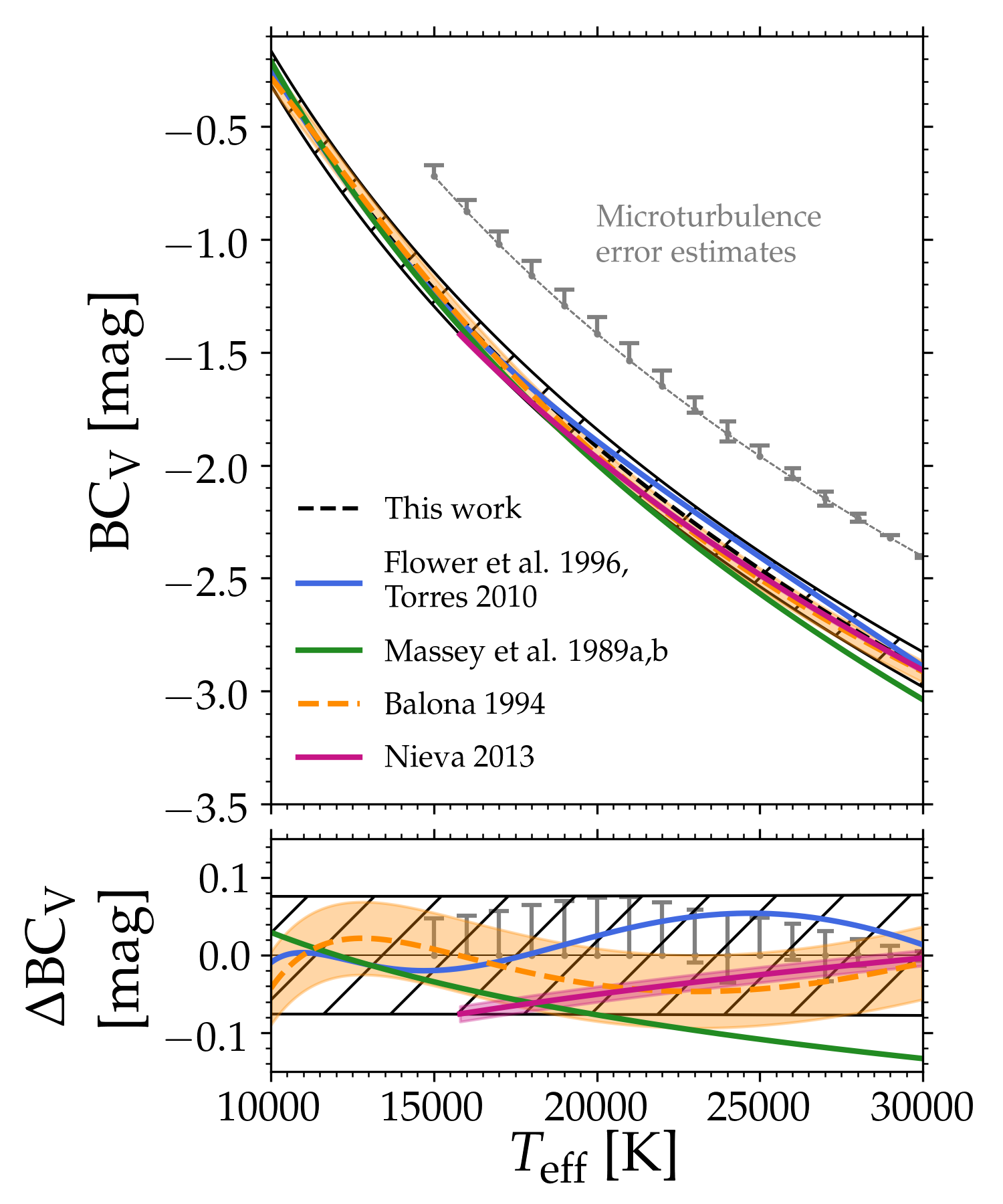}
	\caption{\textcolor{black}{Comparison between bolometric correction prescriptions for the Johnson V band in the literature to the ones derived using Eq.\,(\ref{Eq:Model3}) for the LTE+NLTE grid in this work (dashed black line with the hatched region showing the corresponding error range). The top panel shows the resulting $BC_\text{V}$ vs $T_\text{eff}$ and the bottom panel the corresponding differences. The grey points with uncertainties correspond to the range in the deviation of $BC_\text{V}$ arising from increasing the microturbulence to 10\,km\,s$^{-1}$ at specific $T_\text{eff}$  arbitrarily offset by 0.5\,dex for visibility. $\Delta$BC$_V$ = BC$^\text{Literature}_\text{Johnson.V}$ - BC$^\text{This work}_\text{Johnson.V}$.}}
	\label{fig:BCother}
\end{figure}

\textcolor{black}{We compare our bolometric correction prescription in Eq.\,(\ref{Eq:Model3}) for the LTE+NLTE grid in the Johnson V passband to the literature prescriptions from \citet{Massey1989a,Massey1989b}, \citet{Balona1994}, \citet{Nieva2013}, and \citet{Flower1996} and \citet{Torres2010} in Fig.~\ref{fig:BCother}, for $\log g = 4.0$ and [M/H]=0. The predicted $BC_\text{V}$ values can be quite different depending on the prescription being used. For both \citet{Balona1994} and \citet{Nieva2013} the shaded regions correspond to their 0.047\,mag rms and 0.01\,mag standard deviation, respectively. The grey dashed line with uncertainties at specific $T_\text{eff}$ values has been arbitrarily offset from the black dashed line by 0.5\,dex and show the minimum and maximum deviations in the bolometric corrections $\Delta$BC$^\text{\,turb}_{S_X}$ arising from increasing the microturbulence parameter to 10\,km\,s$^{-1}$ as discussed in Sect.~\ref{Sec:vturb}.} \textcolor{black}{As seen in the bottom panel of Fig.\,\ref{fig:BCother}, the differences in the $BC_\text{V}$ largely fall within our error estimates. While the prescription by \citet{Nieva2013} is also based on NLTE models, it was derived for stars with $T_\text{eff} = 15800-34000$\,K. Because half of the SPB stars considered in this work have $T_\text{eff} < 15800$\,K, we choose not to do any further detailed comparisons for this prescripton and instead focus on the \citet{Flower1996} prescriptions for the remainder of this work. The uncertainties on the bolometric corrections arising from increasing the microturbulence all fall within the uncertainty regions from the error propagation of the regression coefficients.}

\subsection{Comparison to Flower}\label{Sec:BC_flower}

\begin{figure*}
\centering
\begin{minipage}{.33\textwidth}
  \centering
  \includegraphics[width=\linewidth]{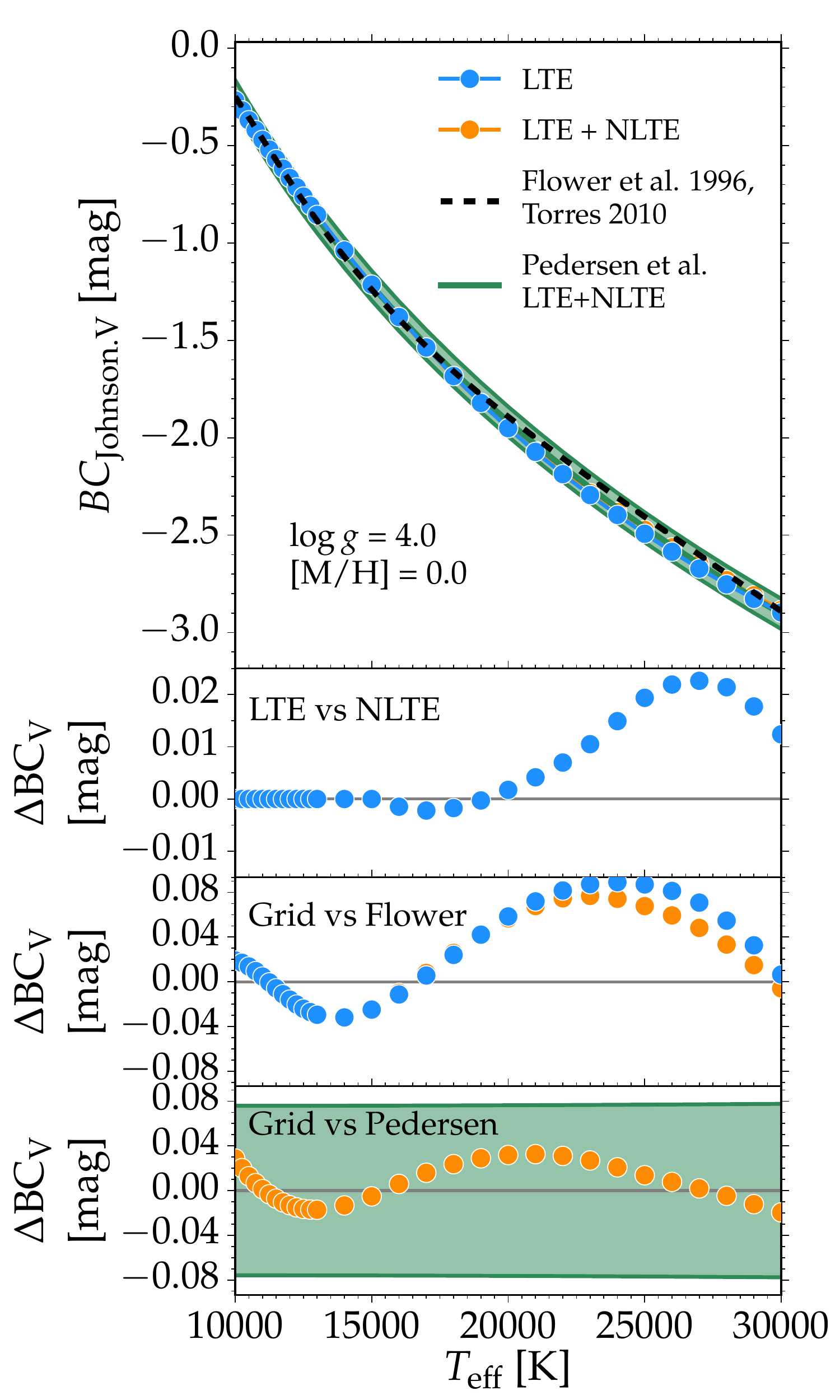}
	%\caption{Calculated bolometric corrections based on Kurucz models for different filters.}
\end{minipage}%
\begin{minipage}{.33\textwidth}
  \centering
  \includegraphics[width=\linewidth]{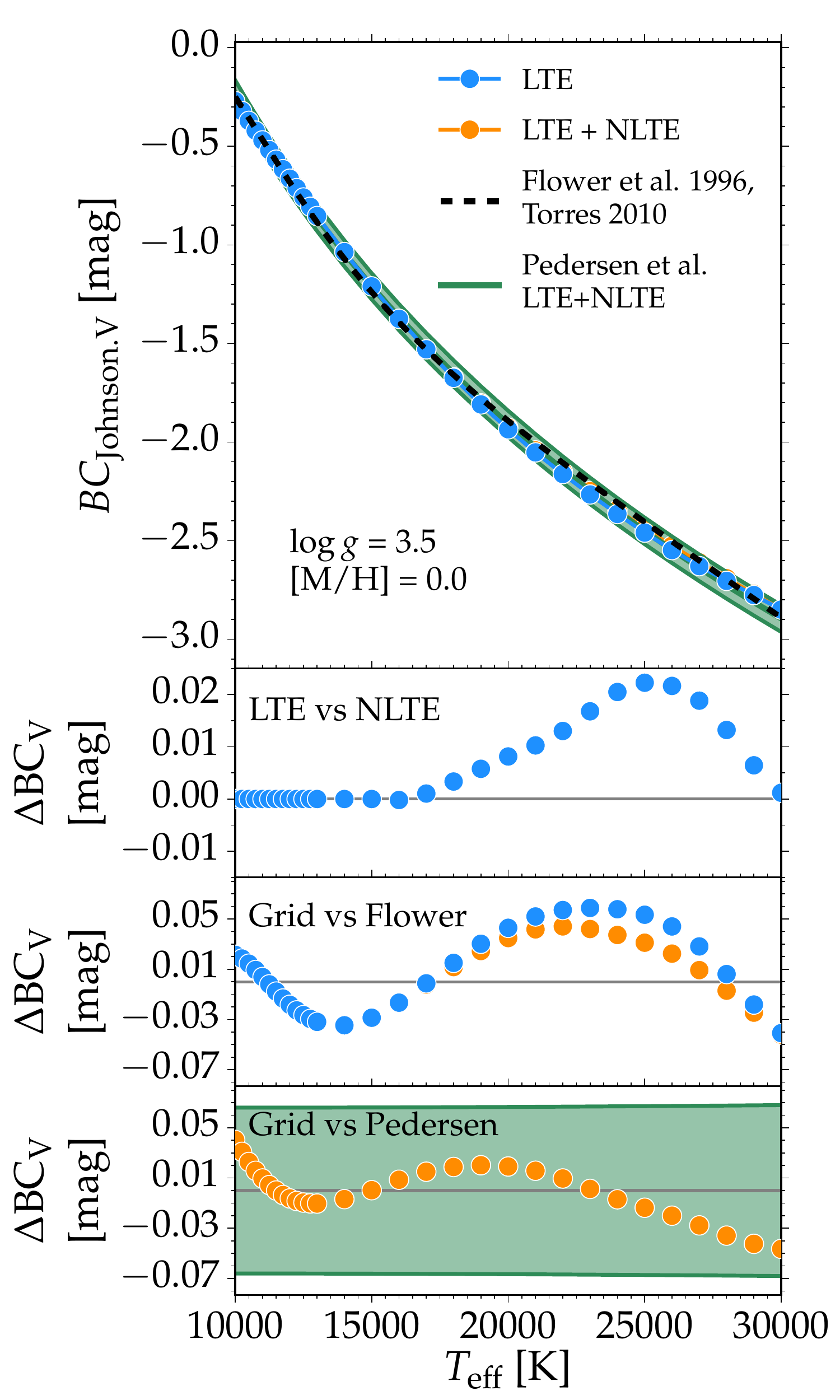}
	%\caption{Same as \emph{left} panel but after adjusting to match solar values.}
\end{minipage}
\begin{minipage}{.33\textwidth}
  \centering
  \includegraphics[width=\linewidth]{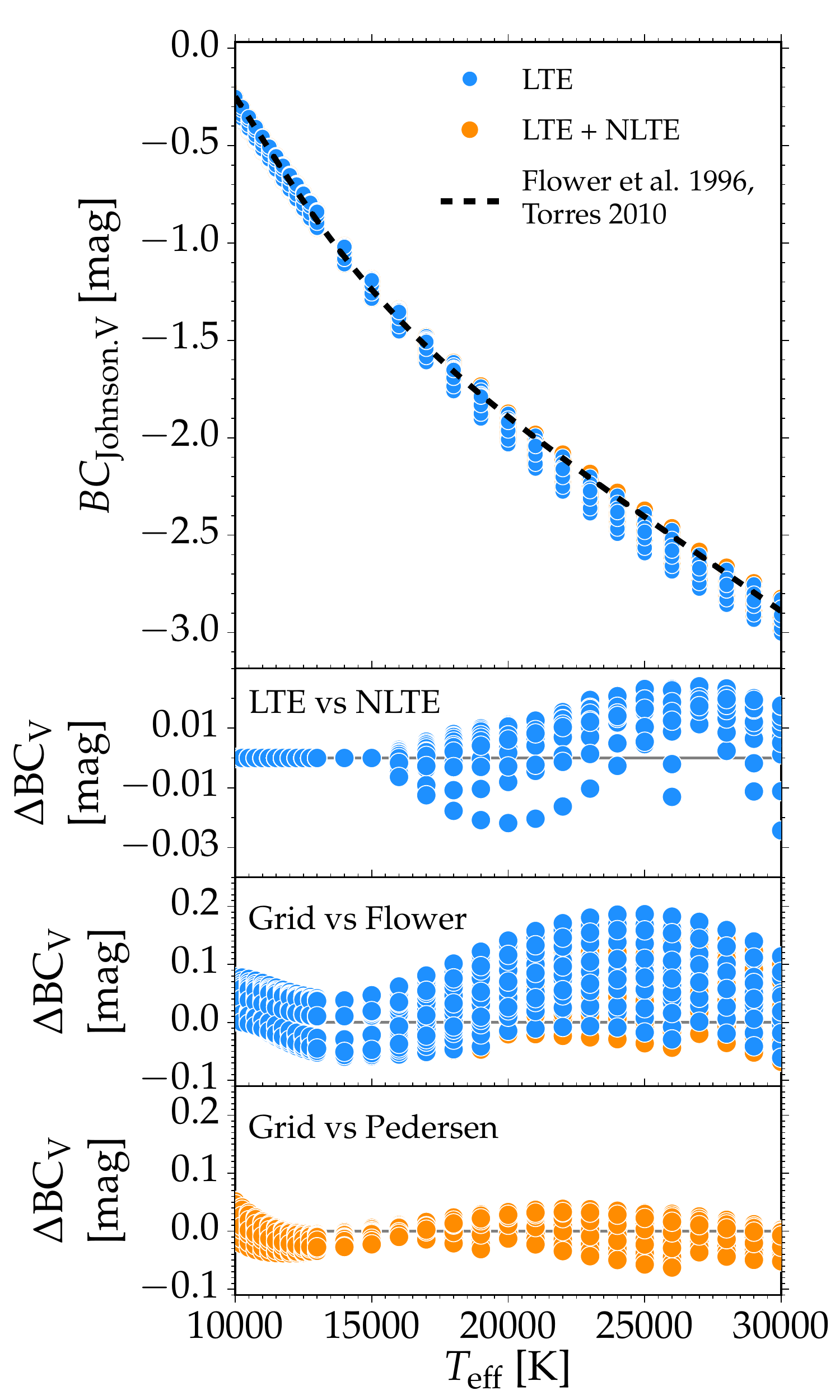}
	%\caption{Same as \emph{left} panel but after adjusting to match solar values.}
\end{minipage}
\caption{Comparison between the bolometric corrections in the Johnson V passband predicted by \citet{Flower1996}, the LTE and LTE+NLTE calculated grid values, and the predictions based on statistical model 3 in Eq.\,(\ref{Eq:Model3}) calculated for the LTE+NLTE grid. For the subfigure on the left and in the center, the $\log g$ and [M/H] have been fixed for the two grids and the statistical model calculated bolometric corrections as annotated in the top panel of the subfigures. All LTE and LTE+NLTE BC$_\text{Johnson.V}$ grid values within 10000--30000\,K are shown in the subfigure on the right. \emph{Top panel:} Black dashed curve show the predicted bolometric corrections by \citet{Flower1996} when using the higher precision coefficients given by \citet{Torres2010}. The green shaded region indicates the model 3 predictions for the fixed $\log g$ and [M/H] values, as is excluded in the figure on the right. The BC$_\text{Johnson.V}$ grid values are shown in blue and orange circles for LTE and LTE+NLTE grid, respectively. \emph{Second panel:} Same as Fig.\,\ref{fig:lte_vs_nlte}. \emph{Third panel:} $\Delta$BC$_V$ = BC$^\text{Flower}_\text{Johnson.V}$ - BC$^\text{grid}_\text{Johnson.V}$. \emph{Fourth panel:} $\Delta$BC$_V$ = BC$^\text{Model 3}_{\text{Johnson.}V}$ - BC$^\text{LTE+NLTE}_{\text{Johnson.}V}$. The green shaded region in the left and center figure shows the $\pm 1\sigma$ standard deviation derived for the statistical model at the given effective temperature.}
	\label{fig:FlowerBC}
\end{figure*}

Figure\,\ref{fig:FlowerBC} illustrates how well the prescription by \citet[][black dashed line]{Flower1996} matches the derived synthetic bolometric corrections in the Johnson\,$V$ passband for both the LTE (blue) and LTE+NLTE (orange) grid. For the construction of the black-dashed line in the top panels, the higher precision values of the coefficients of the \citet{Flower1996} prescription provided by \citet{Torres2010} have been used. For the figure on the left and in the center the results are shown for a fixed value of $\log g$ and [M/H], and also includes in green our derived LTE+NLTE statistical model 3 for these two, fixed parameters. The entire LTE and LTE+NLTE grids in the 10000-30000\,K range have been included in the figure on the right. 

The second panel in each subfigure shows once again the difference between the BC$_{\text{Johnson.}V}$ values of the LTE and LTE+NLTE grid, as also illustrated in Fig.\,\ref{fig:lte_vs_nlte}. The third panel shows the remaining variability in the BC$_{\text{Johnson.}V}$ values for both grids when subtracted from the predictions by \citet{Flower1996}, while the same differences are shown in the fourth panel for our derived statistical model for the LTE+NLTE grid. For the fixed values of [M/H] = 0.0 and $\log g = 4.0$ and 3.5, the differences between the \citet{Flower1996} predictions and the calculated grid values are smaller for the LTE+NLTE grid than the LTE one. \textcolor{black}{The} differences shown in the last panel are generally smaller than those for the \citet{Flower1996} prescription, and all fall within the corresponding error ranges shaded in green. \textcolor{black}{The} bolometric prescriptions predicted by \citet{Flower1996} generally tend to overestimate \textcolor{black}{the} BC$_{\text{Johnson.}V}$ values of the LTE and LTE+NLTE grid by up to 0.2\,dex and that the differences become largest at $\sim 25000$\,K, as shown in the subfigure on the right. In comparison the differences between the predictions by our statistical model 3 and the grid values all stay below $\pm 0.08$\,dex.

%%%%%%%%%%%%%%%%%%%%%%%%%%%%%%%%%%%%%%%%%%%%%%%%%%%%%%%%%%%%%%%%%%%%%%%%%%
%				Luminosities
%%%%%%%%%%%%%%%%%%%%%%%%%%%%%%%%%%%%%%%%%%%%%%%%%%%%%%%%%%%%%%%%%%%%%%%%%%

\begin{table*}
	\centering
	\caption{Distances $d$ determined by \citet{Bailer-Jones2018} and \citet{Anders2019} using the \emph{Gaia} parallaxes $\varpi$, and the corresponding reddening E(B-V) at these distances according to the \citet{Green2019} reddening maps.}
	%\resizebox{\linewidth}{!}{
	\begin{tabular}{@{\extracolsep{4pt}}lccccccc@{}}%@{\hskip 0.3in}
		\hline\\[-1.5ex]
		\text{KIC ID}	&	\text{$\varpi$ [mas]} &	\text{$\varpi_\text{error}$ [\%] }	&	\text{RUWE}	& \multicolumn{2}{c}{\citet{Bailer-Jones2018}}&\multicolumn{2}{c}{\citet{Anders2019}}\\
		\cline{5-6}\cline{7-8}\\[-1.5ex]
        	&	&	&	&	\text{$d$ [kpc]}	&	\text{E(B-V)}&	\text{$d$ [kpc]}	&	\text{E(B-V)}	\\[0.5ex]
        \hline\\[-1.5ex]
		1430353	&	0.079$\pm$0.028	&	\textbf{35}	&	0.97	&	7.760$\pm$1.538	&	0.210$\pm$0.004	&	5.766$\pm$0.916	&	0.210$\pm$0.004\\[0.5ex]
		3240411	&	0.445$\pm$0.060	&	13	&	0.92	&	2.132$\pm$0.292	&	0.100$\pm$0.004	&	1.942$\pm$0.266	&	0.090$\pm$0.002\\[0.5ex]
		3459297	&	0.262$\pm$0.040	&	15	&	1.18	&	3.463$\pm$0.524	&	0.130$\pm$0.003	&	3.009$\pm$0.503	&	0.130$\pm$0.002\\[0.5ex]
		3756031	&	0.499$\pm$0.055	&	11	&	1.20	&	1.915$\pm$0.213	&	0.120$\pm$0.002	&	1.814$\pm$0.214	&	0.120$\pm$0.002\\[0.5ex]
		3839930	&	0.539$\pm$0.055	&	10	&	1.14	&	1.763$\pm$0.178	&	0.090$\pm$0.007	&	1.673$\pm$0.183	&	0.090$\pm$0.007\\[0.5ex]
		3865742	&	0.194$\pm$0.034	&	18	&	0.95	&	4.468$\pm$0.733	&	0.140$\pm$0.002	&	3.618$\pm$0.375	&	0.140$\pm$0.002\\[0.5ex]
		4930889A	&	0.957$\pm$0.047	&	5	&	1.07	&	1.017$\pm$0.050	&	0.090$\pm$0.007	&	0.980$\pm$0.053	&	0.087$\pm$0.007\\[0.5ex]
		4936089	&	0.599$\pm$0.029	&	5	&	1.04	&	1.596$\pm$0.074	&	0.121$\pm$0.014	&	1.532$\pm$0.090	&	0.120$\pm$0.012\\[0.5ex]
		4939281	&	0.215$\pm$0.025	&	11	&	0.98	&	4.103$\pm$0.433	&	0.200$\pm$0.007	&	3.507$\pm$0.370	&	0.190$\pm$0.009\\[0.5ex]
		5309849	&	0.391$\pm$0.032	&	8	&	0.95	&	2.395$\pm$0.189	&	0.250$\pm$0.009	&	2.289$\pm$0.193	&	0.250$\pm$0.009\\[0.5ex]
		6352430A	&	2.607$\pm$0.065	&	2	&	\textbf{1.52}	&	0.380$\pm$0.009	&	0.000$\pm$0.007	&	0.379$\pm$0.011	&	0.000$\pm$0.007\\[0.5ex]
		6462033	&	0.409$\pm$0.033	&	8	&	1.07	&	2.295$\pm$0.178	&	0.254$\pm$0.027	&	2.073$\pm$0.148	&	0.240$\pm$0.021\\[0.5ex]
		6780397	&	0.618$\pm$0.042	&	7	&	1.06	&	1.551$\pm$0.105	&	0.080$\pm$0.007	&	1.476$\pm$0.103	&	0.080$\pm$0.007\\[0.5ex]
		7630417	&	0.046$\pm$0.024	&	\textbf{53}	&	1.03	&	10.346$\pm$2.262	&	0.300$\pm$0.010	&	6.736$\pm$1.645	&	0.300$\pm$0.010\\[0.5ex]
		7760680	&	0.846$\pm$0.050	&	6	&	1.10	&	1.148$\pm$0.068	&	0.100$\pm$0.004	&	1.114$\pm$0.070	&	0.100$\pm$0.004\\[0.5ex]
		8057661	&	0.310$\pm$0.046	&	15	&	1.10	&	2.970$\pm$0.441	&	0.390$\pm$0.019	&	2.644$\pm$0.436	&	0.358$\pm$0.009\\[0.5ex]
		8087269	&	0.397$\pm$0.031	&	8	&	0.99	&	2.336$\pm$0.170	&	0.070$\pm$0.005	&	2.180$\pm$0.211	&	0.070$\pm$0.005\\[0.5ex]
		8255796	&	0.090$\pm$0.020	&	\textbf{23}	&	1.01	&	7.898$\pm$1.251	&	0.390$\pm$0.007	&	5.737$\pm$1.120	&	0.390$\pm$0.007\\[0.5ex]
		8324482	&	0.513$\pm$0.026	&	5	&	0.96	&	1.848$\pm$0.091	&	0.397$\pm$0.008	&	1.759$\pm$0.112	&	0.386$\pm$0.009\\[0.5ex]
		8381949	&	0.217$\pm$0.042	&	19	&	1.02	&	4.019$\pm$0.722	&	0.290$\pm$0.014	&	3.601$\pm$0.586	&	0.290$\pm$0.007\\[0.5ex]
		8459899	&	1.375$\pm$0.126	&	9	&	\textbf{2.35	}&	0.722$\pm$0.069	&	0.140$\pm$0.024	&	0.702$\pm$0.075	&	0.139$\pm$0.025\\[0.5ex]
		8714886	&	0.615$\pm$0.041	&	7	&	1.08	&	1.560$\pm$0.102	&	0.347$\pm$0.014	&	1.505$\pm$0.109	&	0.341$\pm$0.012\\[0.5ex]
		8766405	&	0.691$\pm$0.041	&	6	&	1.04	&	1.393$\pm$0.082	&	0.150$\pm$0.001	&	1.296$\pm$0.082	&	0.150$\pm$0.007\\[0.5ex]
		9020774	&	0.099$\pm$0.032	&	\textbf{32}	&	1.01	&	6.355$\pm$1.145	&	0.070$\pm$0.007	&	6.114$\pm$1.680	&	0.070$\pm$0.007\\[0.5ex]
		9227988	&	0.095$\pm$0.033	&	\textbf{35}	&	1.04	&	6.799$\pm$1.408	&	0.170$\pm$0.004	&	5.126$\pm$1.107	&	0.170$\pm$0.004\\[0.5ex]
		9964614	&	0.262$\pm$0.041	&	16	&	1.10	&	3.407$\pm$0.494	&	0.110$\pm$0.005	&	3.005$\pm$0.355	&	0.110$\pm$0.005\\[0.5ex]
		9715425	&	0.100$\pm$0.041	&	\textbf{41}	&	1.13	&	6.059$\pm$1.324	&	0.120$\pm$0.010	&	4.524$\pm$1.115	&	0.120$\pm$0.010\\[0.5ex]
		10285114	&	0.485$\pm$0.035	&	7	&	0.99	&	1.948$\pm$0.135	&	0.070$\pm$0.007	&	1.847$\pm$0.154	&	0.070$\pm$0.016\\[0.5ex]
		10526294	&	0.297$\pm$0.030	&	10	&	0.98	&	3.025$\pm$0.274	&	0.060$\pm$0.002	&	2.733$\pm$0.337	&	0.060$\pm$0.002\\[0.5ex]
		10536147	&	0.050$\pm$0.044	&	\textbf{87}	&	0.98	&	6.776$\pm$1.481	&	0.090$\pm$0.009	&	5.524$\pm$1.004	&	0.090$\pm$0.009\\[0.5ex]
		10658302	&	0.152$\pm$0.041	&	\textbf{27}	&	1.08	&	4.733$\pm$0.853	&	0.070$\pm$0.007	&	4.215$\pm$1.001	&	0.070$\pm$0.007\\[0.5ex]
		11360704	&	0.258$\pm$0.039	&	15	&	1.05	&	3.428$\pm$0.476	&	0.120$\pm$0.007	&	2.973$\pm$0.359	&	0.120$\pm$0.007\\[0.5ex]
		11971405	&	1.000$\pm$0.036	&	4	&	0.94	&	0.973$\pm$0.035	&	0.095$\pm$0.014	&	0.953$\pm$0.038	&	0.092$\pm$0.014\\[0.5ex]
		12258330	&	1.064$\pm$0.040	&	4	&	1.10	&	0.916$\pm$0.034	&	0.070$\pm$0.012	&	0.897$\pm$0.040	&	0.070$\pm$0.012\\[0.5ex]
        \hline
        \end{tabular}
        \vspace{1ex}\\
	
	\raggedright\small \textbf{Notes:} RUWE = re-normalised unit weight error, discussed in Sect.\,\ref{Sec:Binaries}. Values larger than 1.40 are marked in bold characters. The same is done for the parallax errors labeled in procent for $\varpi_\text{error} > 20$ per cent.\\
        \label{Tab:Distances_EBVs}
\end{table*}

%\clearpage
\section{Derivation of luminosities}\label{Sec:Lum_derivarion}

The derivation of luminosities from measured apparent magnitudes relies on knowing a) the appropriate bolometric corrections $\text{BC}_{S_\lambda}$ to be used in the conversion to bolometric magnitudes, b) the distance $d$ to the star, and c) the interstellar line-of-sight extinction $A_{S_\lambda}$\textcolor{black}{.}

\subsection{Luminosities from apparent magnitudes and bolometric corrections}\label{Sec:Lum_from_BC}

Rewriting Eq.\,(\ref{Eq:Lum_BC_Mfilter}), the luminosities $L_\star$ are deduced from measured photometric magnitudes using

\begin{align}
-2.5 \log L_\star/ L_\odot %&= M_\text{bol} - M_{\text{bol},\odot}\nonumber\\
	&= m_\text{bol} - 5 \log d + 5 - M_{\text{bol},\odot}.
	\label{Eq:Lum}
\end{align}

\noindent \textcolor{black}{For the distances $d$ we consider both the ones derived by \citet{Bailer-Jones2018} and \citet{Anders2019} based on \emph{Gaia} DR2 parallaxes.} The parameters $M_\text{bol}$ and $m_\text{bol}$ denote the extinction corrected absolute and apparent bolometric magnitudes of the star. The apparent bolometric magnitude is obtained by applying the bolometric correction $BC_{S_\lambda}$ to the apparent magnitude $m_{S_\lambda}$ measured in a photometric passband with transmission curve $S_\lambda$:

\begin{equation}
m_\text{bol} = m_{S_\lambda} + BC_{S_\lambda} - A_{S_\lambda},
	\label{Eq:mbol}
\end{equation}

\noindent with $A_{S_\lambda}$ the extinction in the passband $S_\lambda$. In order to circumvent any uncertainties arising from the choice of measured magnitude, $m_\text{bol}$ is calculated for all available $m_{S_\lambda}$ listed independently for each star in Table\,E1 in Appendix\,E \textcolor{black}{available online}. The corresponding bolometric corrections are deduced based on the spectroscopic, \emph{Gaia}, and KIC\footnote{For the KIC parameters we assume errors of $\pm 1000$\,K, 0.5\,dex and 0.1\,dex of $T_\text{eff}$, $\log g$, and [M/H], respectively, for the calculation of the bolometric corrections.} stellar parameters given in Table\,F1 in Appendix~F \textcolor{black}{available online}, with the $\text{BC}_\text{Gaia.BPfaint}$ being applied for stars with $G > 10.86$ and $\text{BC}_\text{Gaia.BPbright}$ for $G \leq 10.86$ \citep{MaizApellanizWeiler2018}. The final value of the apparent bolometric magnitude is then taken as the weighted average of all computed $m_\text{bol}$ for the star, while the standard deviation provides the corresponding error.

In order to derive extinction corrected values of $m_\text{bol}$, the extinction $A_{S_\lambda}$ is deduced using an (observed) reddening $E(B-V)$ of the star and a reddening law for the ratio of total to selective extinction $R_{S_\lambda}$:

\begin{equation}
A_{S_\lambda} = R_{S_\lambda}  E(B-V).
	\label{Eq:A_extinction}
\end{equation}

\noindent The $R_{S_\lambda}$ values are determined from the $R_{\lambda}$ vs $\lambda$ curve assuming the \citet{Fitzpatrick2004} reddening law and $R_V = 3.1$ for Vega, see Appendix\,C \textcolor{black}{available online}, while $E(B-V)$ is obtained using the 3D reddening maps of \citet{Green2018,Green2019}. These maps were determined using high-quality Pan-STARRS 1 and 2MASS photometry of 800 million stars (\textsf{Bayestar17} reddening map). An updated version of the map (\textsf{Bayestar19}) has been released \citep{Green2019}, and included also the \emph{Gaia} DR2 parallaxes and photom\textcolor{black}{e}try and the ALLWISE photometry in the calculation of the map. The reddening E(B-V) values from \textsf{Bayestar19} are included for the 34 SPB stars in Table\,\ref{Tab:Distances_EBVs} for both considered distances. A comparison between the \textsf{Bayestar17} and \textsf{Bayestar19} reddening maps for the \emph{Kepler} field-of-view are shown in \textcolor{black}{Fig.\,4.7 in \citet{Pedersen2020PhDT}} at three different distances. The interstellar reddening is extracted from these maps using the \textsf{dustmap} \textsc{python} package \citep{Green2018a}.

\begin{figure}%[h]
  \centering
  \includegraphics[width=0.95\linewidth]{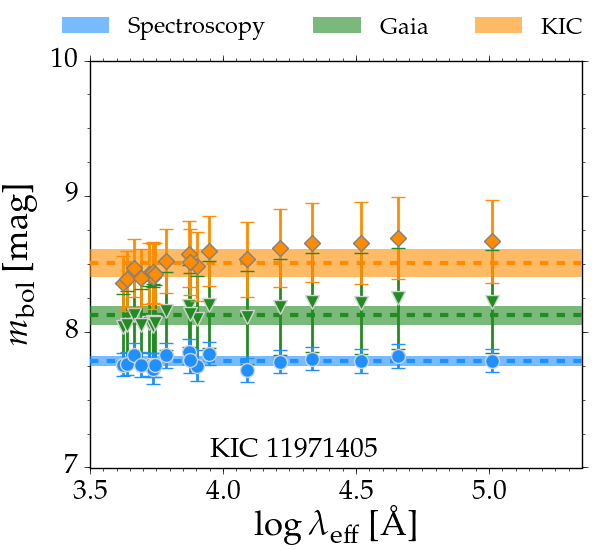}
\caption{Extinction corrected apparent bolometric magnitudes for KIC\,11971405 calculated using Model 3 in Eq.\,(\ref{Eq:Model3}) for the bolometric corrections, and the spectroscopic (blue), \emph{Gaia} (green), and KIC (orange) stellar parameters listed in Table\,F1. The apparent bolometric magnitudes determined for each filter are shown as a function of the effective wavelength of the filter\textcolor{black}{, cf. Appendix~B available online}. The central dashed line indicates the final, determined weighted average of the apparent bolometric correction, and the shaded region the corresponding standard deviation.}
	\label{fig:mbol}
\end{figure}

An example of the resulting extinction-corrected apparent bolometric magnitudes obtained using Eq.\,(\ref{Eq:mbol}) is shown in Fig.\,\ref{fig:mbol} for each filter as a function of its effective wavelength. For this example, the statistical Model 3 representation of $B_{S_\lambda}$ has been used for the derivation of the bolometric corrections. The central dashed line shows the computed weighted average of $m_\text{bol}$ and the shaded regions its standard deviation. 

The resulting $m_\text{bol}$ values should be independent of $\lambda_\text{eff}$ and fall on a straight line. As seen in Fig.\,\ref{fig:mbol}, this is generally the case when the spectroscopic parameters are used for calculating the apparent bolometric corrections in this given example. However, when the \emph{Gaia} parameters from \citet{Anders2019} and the KIC parameters are used, a tilt is introduced in the $m_\text{bol}$ vs $\lambda_\text{eff}$ data, with the apparent bolometric magnitudes increasing towards longer wavelengths. Furthermore, the $m_\text{bol}$ increases on average for both the \emph{Gaia} and KIC parameters, with the increase being larger for the KIC values for which the discrepancy between the effective temperature is largest ($\sim$3500\,K smaller than the value from spectroscopy). This in turn, would lead to an underestimation of the final derived luminosity and illustrates the importance of having well determined effective temperatures when deriving stellar luminosities. 

\textcolor{black}{The stellar parameters from the \emph{Kepler} input catalog and the corresponding \emph{Gaia} parameters from \citet{Anders2019} are derived from photometric data. While the approaches adopted for the parameters in the KIC and by \citet{Anders2019} work well for late type stars, the issue that arises for stars with increasing effective temperatures is that the SED moves towards shorter wavelengths no longer covered by the photometric passbands, cf. Fig.~A1. As a consequence, it is very difficult to obtain reliable stellar temperatures based on photometry alone for stars with $T_\text{eff} > 10000$\,K and one should use $m_\text{bol}$ values deduced from spectroscopy in this temperature regime.}

\textcolor{black}{A} decrease in $m_\text{bol}$ can \textcolor{black}{also} arise from the re-emission of absorbed stellar light by circumstellar material, which increases the brightness of the star at longer wavelengths. Our methodology for determining $m_\text{bol}$ does not take such re-emission into account\textcolor{black}{.} Therefore, \textcolor{black}{we exclude} the WISE data from the luminosity calculations when such discrepancies are seen in the calculated apparent bolometric corrections.

\subsection{Luminosities for binary systems}\label{Sec:Binaries}

The re-normalised unit weight error (RUWE) defined by \citet{Lindegren2018a,Lindegren2018b}\footnote{\url{https://www.cosmos.esa.int/documents/29201/1770596/Lindegren\_GaiaDR2\_Astrometry\_extended.pdf/1ebddb25-f010-6437-cb14-0e360e2d9f09}}, is a goodness-of-fit indicator for how reliable the \emph{Gaia} astrometric data is for a given target. \citet{Lindegren2018b} found that for 70 per cent of the best \emph{Gaia} sources, the RUWE falls below 1.40. Therefore, this value has generally been taken as a cut-off for when the \emph{Gaia} astrometry is considered reliable, and \citet{Anders2019} likewise flag the \emph{Gaia} targets according to if their RUWE is smaller or larger than 1.40. The RUWE values calculated by \citet{Anders2019} based on \citet{Lindegren2018b} are listed in Table\,\ref{Tab:Distances_EBVs} for the \textcolor{black}{SPB} stars considered in this work.

A likely cause of sources having $\text{RUWE} > 1.40$, is that the stars are in fact in binary systems. In the case of unresolved binaries with separations smaller than 100\,mas, both orbital and photometric variability may give rise to biases in the astrometric parameters \citep{Lindegren2018a}. For partially and fully resolved binaries, the changes in the direction of the scanned region of the sky may cause changes to which of the components in the binary system that the observations are carried out for, while the source is labeled as being the same. This \textcolor{black}{impacts} the derived parallaxes \citep{Lindegren2018a,Arenou2018}. For binaries consisting of `twin' systems where the components are (close to) identical, the astrometry is expected to still be reliable because the photocenter of the system behaves like a single star\footnote{\url{https://www.arcetri.inaf.it/~mathieu/.EwAsS-2019-SS22/301_ss22a_0900_pourbaix.pdf}}.

Out of the 34 SPB stars considered in this work, two are known and confirmed spectroscopic binaries. KIC\,4930889 is a system consisting of a B5IV- and B8IV-V star, and has an orbital period of 18.296$\pm$0.002\,d \citep{Papics2017}. In comparison the other known system KIC\,6352430  has a period of 26.551$\pm$0.019\,d, and consists of a B7V and F2.5V star \citep{Papics2013}. \textcolor{black}{Only} KIC\,6352430 is marked as having $\text{RUWE} > 1.40$, while $\text{RUWE} = 1.07$ for KIC\,4930889. This could be because the two components of KIC~4930889 are relatively close in spectral type. The stars with the highest RUWE value ($\text{RUWE} = 2.35$) is KIC\,8459899. \citet{Lehmann2011} suspected this star to be a double-lined spectroscopic \textcolor{black}{binary.} The possible binary nature of the star is backed up by its very high RUWE value. On the other hand the RUWE value alone is not enough for identifying binary systems in the \emph{Gaia} data as shown for KIC\,4930889. For all other stars listed in Table\,\ref{Tab:Distances_EBVs}, $\text{RUWE} < 1.40$.

In spite of the possibly inaccurate parallaxes and distances, \textcolor{black}{we calculate} the luminosities of these \textcolor{black}{three} systems using the spectroscopic parameters of the primary components and consider these as first estimates of the actual luminosities. For KIC\,6352430 the large difference in spectral types between the two components means that the light contribution from the secondary is \textcolor{black}{minimal.} In the case of KIC\,493088, 71 per cent of the light comes from the primary component \citep{Papics2017}. For KIC\,8459899, the spectral type of the secondary component is unknown and \textcolor{black}{cannot} be corrected for. Fully reliable luminosities for \textcolor{black}{these binaries} will have to await the \emph{Gaia} data release \textcolor{black}{3.}

\textcolor{black}{A binary fraction of 9\% among the SPB stars considered in this work may seem low in comparison to the expected $25 \pm 2$\,\% ($58 \pm 11$\,\%) estimated by \citet{Dunstall2015} for B-type stars before (after) correcting for observational biases. This low detection rate is due to a lack of multi-epoch spectroscopic observations with good radial velocity estimates. Eight of the stars listed in Table~F1 have low resolution (R=1800) LAMOST spectra available covering multiple epochs, however the high uncertainties on the radial velocity measurements \citep[$20-30$\,km\,s$^{-1}$][]{Frasca2016} makes it impossible to distinguish binary systems from single stars for these eight SPBs.}

\subsection{Comparison of luminosities}\label{Sec:Lum_comparison}

Table\,G1-G3 \textcolor{black}{in Appendix~G available online} list the luminosities derived in this work for the 34 SPB stars using a variety of different setups. As a general standard, we take the luminosities derived using the spectroscopic parameters, Model 3 from Eq.\,(\ref{Eq:Model3}) based on the LTE+NLTE bolometric correction grid, the \emph{Gaia} distances by \citet{Anders2019}, and the \textsf{Bayestar19} reddening maps from \citet{Green2019}. These luminosities are labeled as $L_\star$ in Table\,G1. All the derived luminosites are compared against these values, unless otherwise specified.

\subsubsection{LTE vs LTE+NLTE}

\begin{figure*}%[h]
  \centering
  \includegraphics[width=\linewidth]{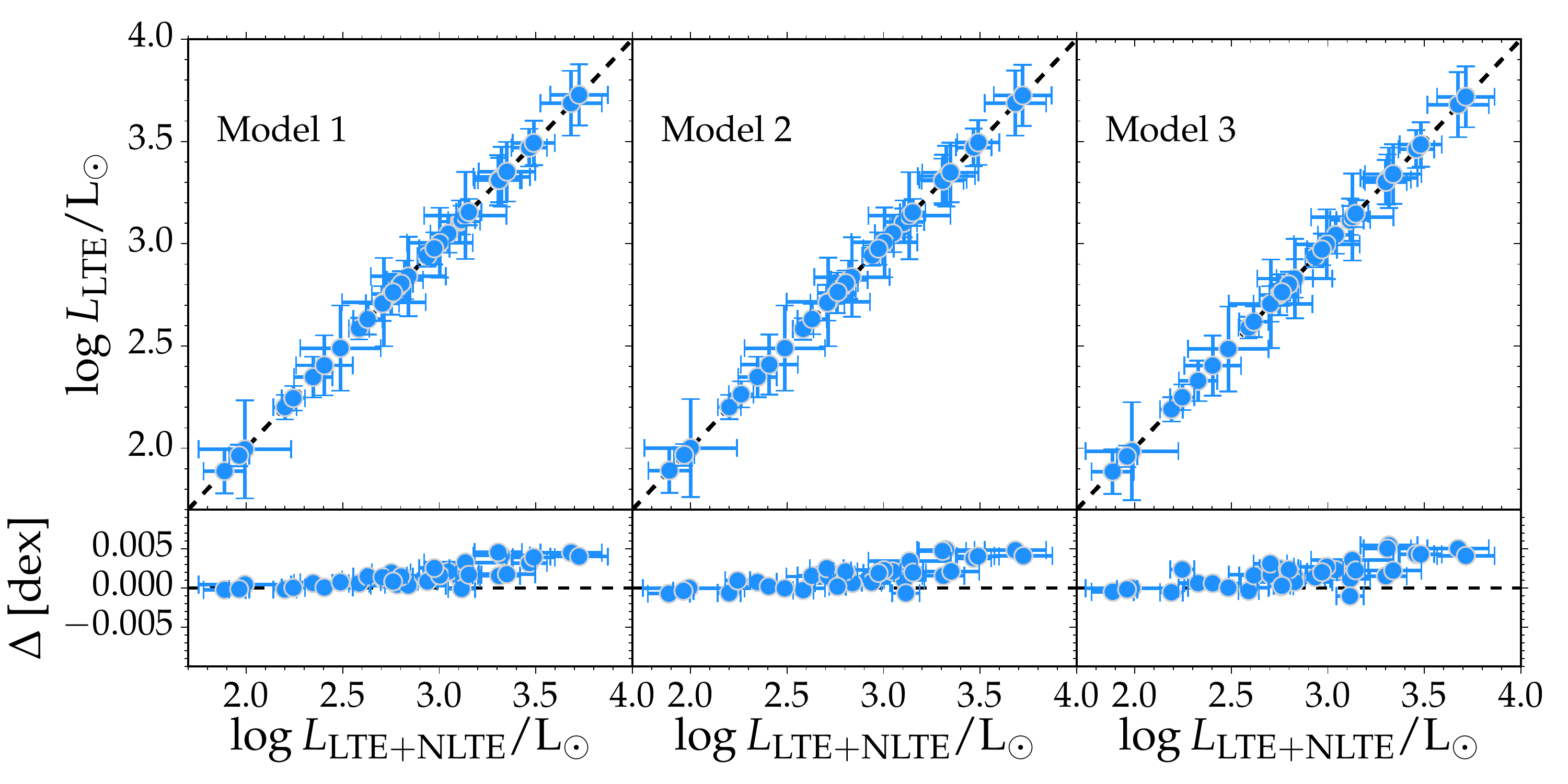}
\caption{Comparison of derived luminosities using the statistical model 1, 2, and 3 (left, center, and right panel) representation of $\text{BC}_{S_\lambda}$ based on the LTE (y-axis on top panels) and LTE+NLTE grid (x-axis). The black dashed lines shows the position for $L_\text{LTE} = L_\text{LTE+NLTE}$, and the bottom panels the residuals. \textcolor{black}{The plotted data are listed in Table~G1.} $\Delta = L_\text{LTE} - L_\text{LTE+NLTE}$.}
	\label{fig:Lum_lte_vs_nlte}
\end{figure*}

Figure\,\ref{fig:Lum_lte_vs_nlte} shows the differences in derived luminosities when the LTE grid is used to determine the statistical model representation of $B_{S_\lambda}$ ($L_\text{LTE}$) instead of the LTE+NLTE grid ($L_\text{NLTE+LTE}$). The top panels show the luminosities based on the LTE grid as a function of $L_\text{NLTE+LTE}$ using Model 1, 2, and 3, respectively, and the bottom panels the corresponding residuals $\Delta$. The errors on $\Delta$ have been omitted in this figure for the sake of clarity. The black-dashed line shows where $L_\text{LTE}= L_\text{NLTE+LTE}$ and $\Delta = 0$ for the top and bottom panels, respectively. In general, a small increase is seen for $L_\text{LTE}$ at higher $L_\text{NLTE+LTE}$ values, but the differences never become larger than \textcolor{black}{$0.01$}\,dex. This is well within the errors on the determined luminosities, and hence we conclude that the impact of using LTE models instead of NLTE is insignificant for the temperature range \textcolor{black}{11500-21000\,K}\textcolor{black}{.}

\subsubsection{Statistical models vs linear interpolation}

\begin{figure*}
\centering
\begin{minipage}{.5\textwidth}
  \centering
  \includegraphics[width=0.9\linewidth]{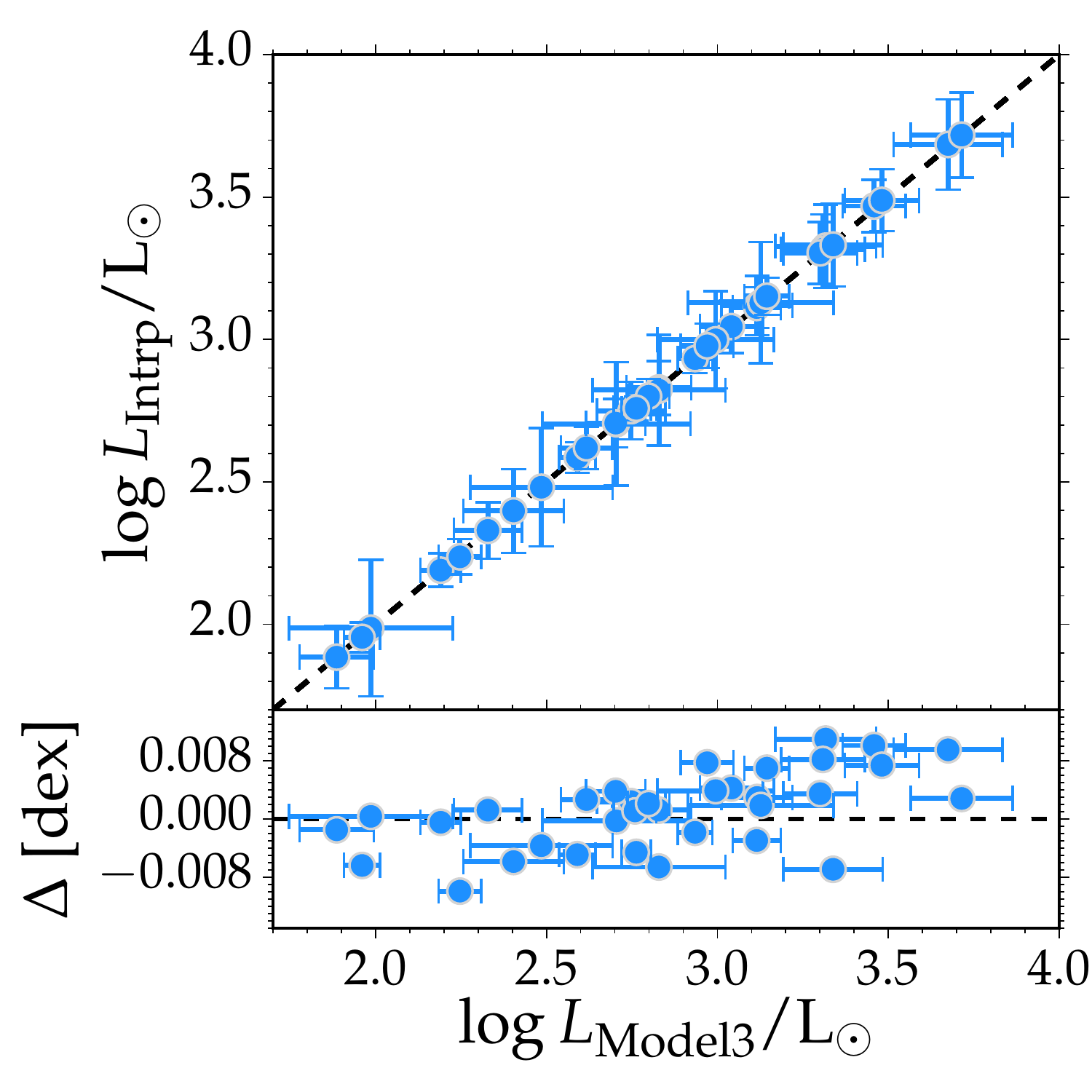}
	%\caption{Calculated bolometric corrections based on Kurucz models for different filters.}
\end{minipage}%
\begin{minipage}{.5\textwidth}
  \centering
  \includegraphics[width=0.9\linewidth]{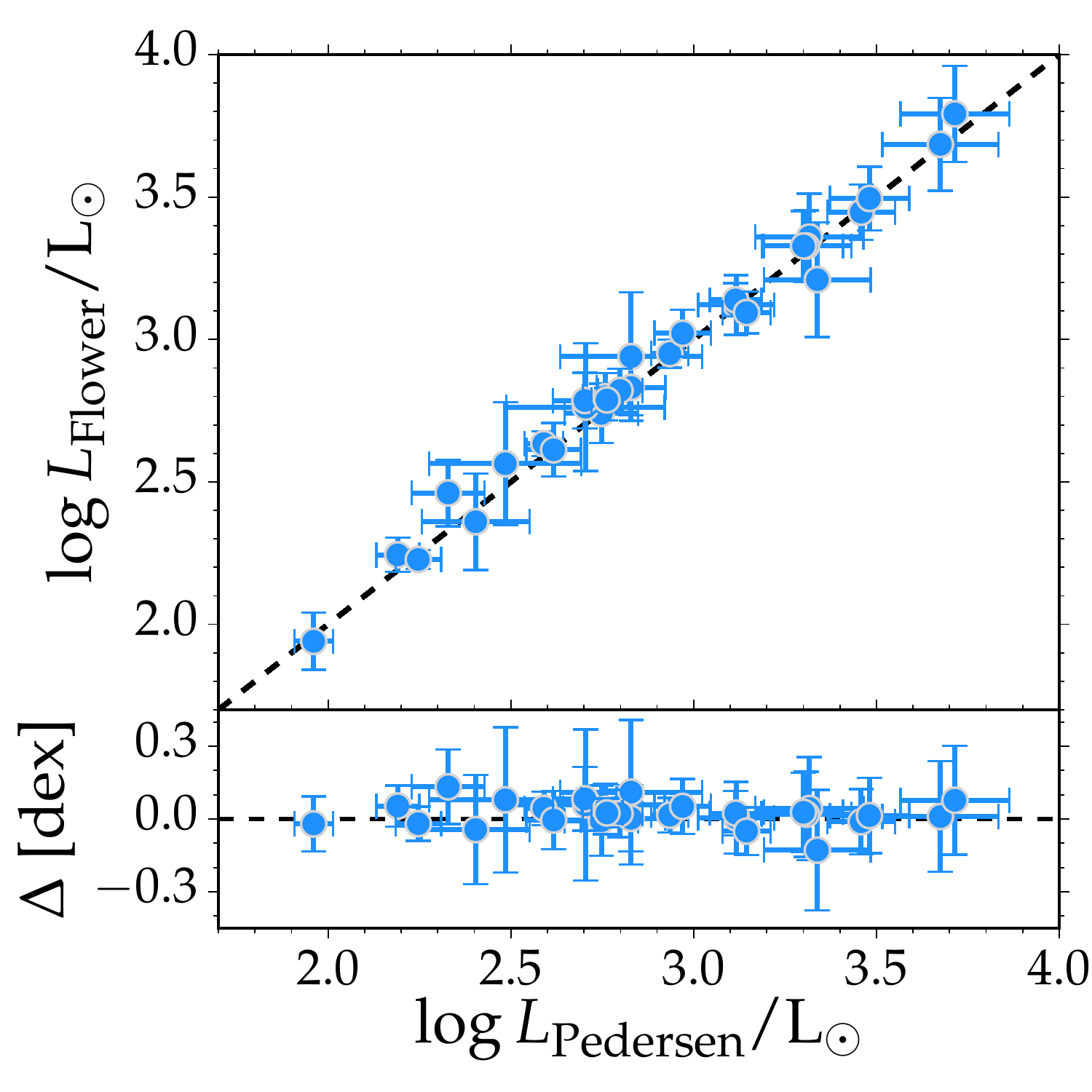}
	%\caption{Same as \emph{left} panel but after adjusting to match solar values.}
\end{minipage}
\caption{\textcolor{black}{\emph{Left:}} Comparison of derived luminosities using a linear interpolation onto the LTE+NLTE $\text{BC}_{S_\lambda}$ grid (\textcolor{black}{y-axis}) and the statistical model\,3 (\textcolor{black}{x-axis}). The errors on the residuals have been omitted in the bottom panel for the sake of clarity. \textcolor{black}{The plotted data on the y-axis are listed in the second column of Table~G2.} \textcolor{black}{\emph{Right:} Comparison between the luminosities} derived using the bolometric correction prescription by \citet{Flower1996} with the updated coefficients from \citet{Torres2010} \textcolor{black}{and the corresponding luminosities derived using our statistical model 3. Both are based on the JOHNSON.V magnitudes, hence o}nly stars with measured V-band magnitudes are included in this subfigure. \textcolor{black}{The plotted data are listed in the fourth and last column of Table~G2.}}
	\label{fig:Lum_comp1}
\end{figure*}

Instead of using a statistical model representation of the bolometric correction\textcolor{black}{, one can choose} to obtain $BC_{S_\lambda}$ through a multidimensional linear interpolation of the observed $T_\text{eff}$, $\log g$, and [M/H] onto the grid. The errors on the interpolated bolometric corrections can then be obtained by repeating the interpolation 1000 times, each time using a different $T_\text{eff}$, $\log g$ and [M/H] drawn from a normal distribution centered around the observed values and using their $1\sigma$ errors as the standard deviation. The mean and standard deviation of the resulting distribution are then taken as $BC_{S_\lambda}$ and its error for the star in a given filter. The differences in the final derived luminosities using this approach and using Model 3 for the bolometric corrections is shown in the left panel of Fig.\,\ref{fig:Lum_comp1}. \textcolor{black}{We see that} the luminosities based on the interpolated bolometric corrections are smaller at \textcolor{black}{lower} values of $L_\text{Model 3}$, and become larger than the statistical model values at increasing luminosities. In all cases, the differences stay below 0.01\,dex and are generally smaller than that.

\subsubsection{Statistical model 3 vs Flower}

The right panel of Fig.\,\ref{fig:Lum_comp1} shows the differences in derived luminosities when the prescription by \citet{Flower1996} is used for the derivation of the bolometric corrections. For these calculations, the higher precision version of the coefficients presented by \citet{Torres2010} are used, and the corresponding luminosity $L_\text{Flower}$ is determined solely from the measured JOHNSON.V magnitudes listed in Table\,E1. \textcolor{black}{The luminosities $L_\text{Pedersen}$ rely also only on the JOHNSON.V magnitudes but are otherwise derived the same way as $L_\star$. Both the $L_\text{Flower}$ and $L_\text{Pedersen}$ values are listed in Table~G3.}
The derived luminosities are scattered around the black, dashed, $L_\text{Flower} = L_\text{Pedersen}$ line in the right panel of Fig.\,\ref{fig:Lum_comp1}, and no clear trend is seen in the residuals. The differences range between \textcolor{black}{0.00}--0.13\,dex, and are generally covered by the errors.

\subsubsection{Spectroscopy, \emph{Gaia}, and KIC parameters}

\begin{figure*}%[h]
  \centering
  \includegraphics[width=\linewidth]{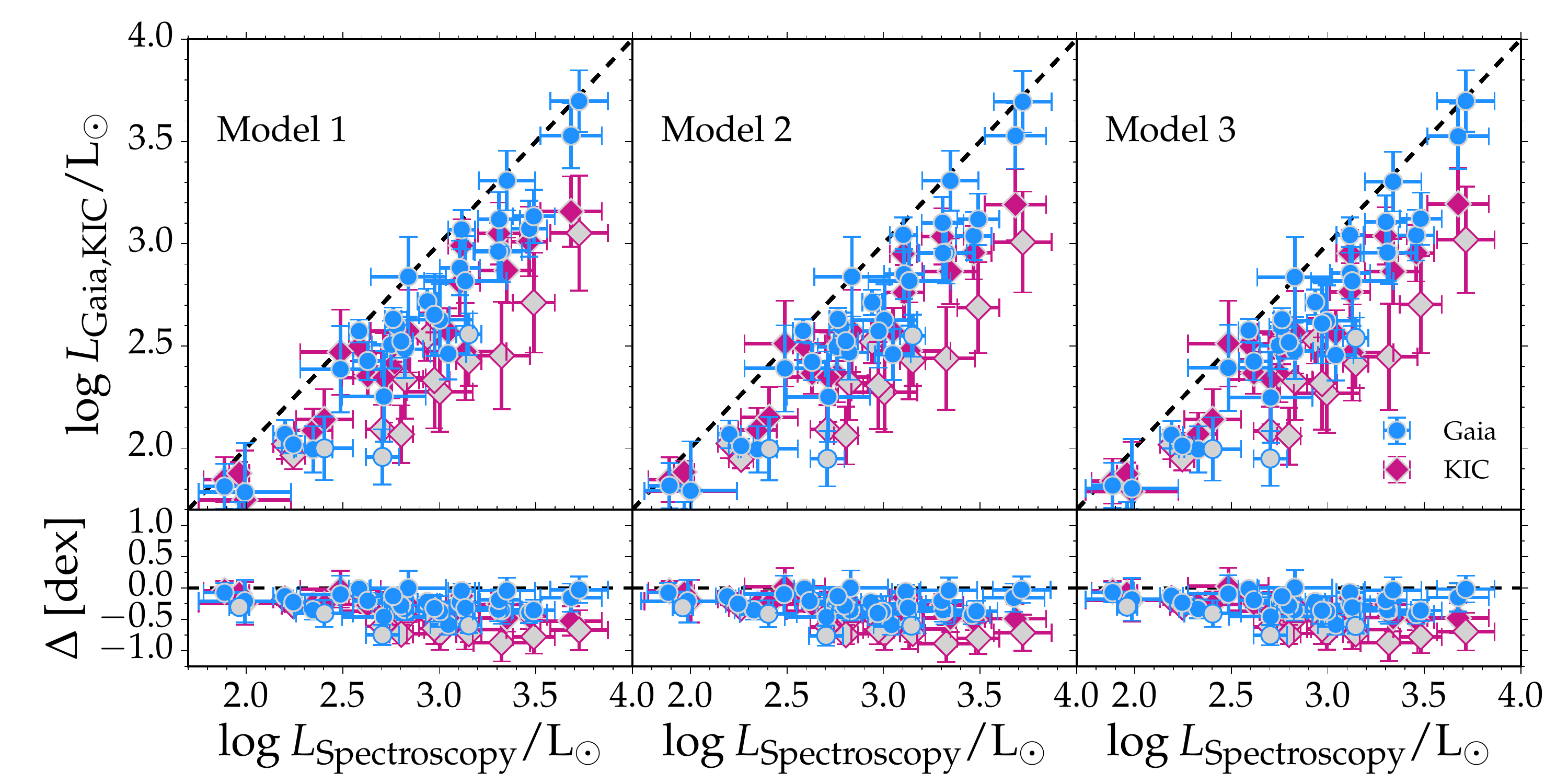}
\caption{Comparison between the luminosities derived based on spectroscopic parameters (all x-axes), and those derived based on the \emph{Gaia} (blue circles) and KIC (pink diamonds) parameters, respectively. For the left, center, and right panels the statistical models 1, 2, and 3 have been used in the derivation of the bolometric corrections, respectively. The residuals are shown in the bottom panels. Black dashed lines in the top panels show the expected position for equal luminosities, and $\Delta = 0$ in the bottom panels. \textcolor{black}{Symbols for which the colours are inverted (i.e. grey centers with blue or pink edges) correspond to stars for which the \emph{Gaia} or KIC effective temperatures are lower than $10000$\,K. The plotted data on the y-axes are listed in Table~G3.}}
	\label{fig:Lum_spec_vs_gaia_vs_kic}
\end{figure*}

The choice of stellar parameters is \textcolor{black}{important} for the derived luminosities \textcolor{black}{(see Fig.\,\ref{fig:mbol})}. This is illustrated \textcolor{black}{in} Fig.\,\ref{fig:Lum_spec_vs_gaia_vs_kic} which compares the luminosities derived from spectroscopy (x-axis) to those based on the \emph{Gaia} and KIC parameters. In the vast majority of the cases, the luminosities are significantly underestimated when either the \emph{Gaia} or KIC parameters are used\textcolor{black}{. The} KIC parameters are known to be inaccurate for hot stars with $T_\text{eff} > 10000$\,K, which has been taken as the cut-off value for that catalogue's temperature estimation. \textcolor{black}{The KIC $T_\text{eff}$ values range from 8150-14800\,K with a median of 10900\,K, whereas the effective temperatures from \citet{Anders2019} range from 8000-19700\,K with a median of 12200\,K. In comparison, the spectroscopic values range from 11500-21000\,K with a median of 16150\,K.} \textcolor{black}{The stars for which the \emph{Gaia} or KIC $T_\text{eff}$ values fall outside of the validity range of the statistical models are marked by the inverted colour symbols in Fig.~\ref{fig:Lum_spec_vs_gaia_vs_kic}.} \textcolor{black}{Excluding these stars,} we find that the discrepancies vary from \textcolor{black}{0.00-0.59}\,dex for the \emph{Gaia} luminosities, and \textcolor{black}{0.02-0.66}\,dex for the KIC luminosities independent on the statistical model being used. Therefore, we conclude that spectroscopic $T_\text{eff}$ are required in order to obtain meaningful luminosities of B-type stars.

%\clearpage
\subsubsection{Bailer-Jones vs Anders}

\begin{figure}
\centering
  \includegraphics[width=0.9\linewidth]{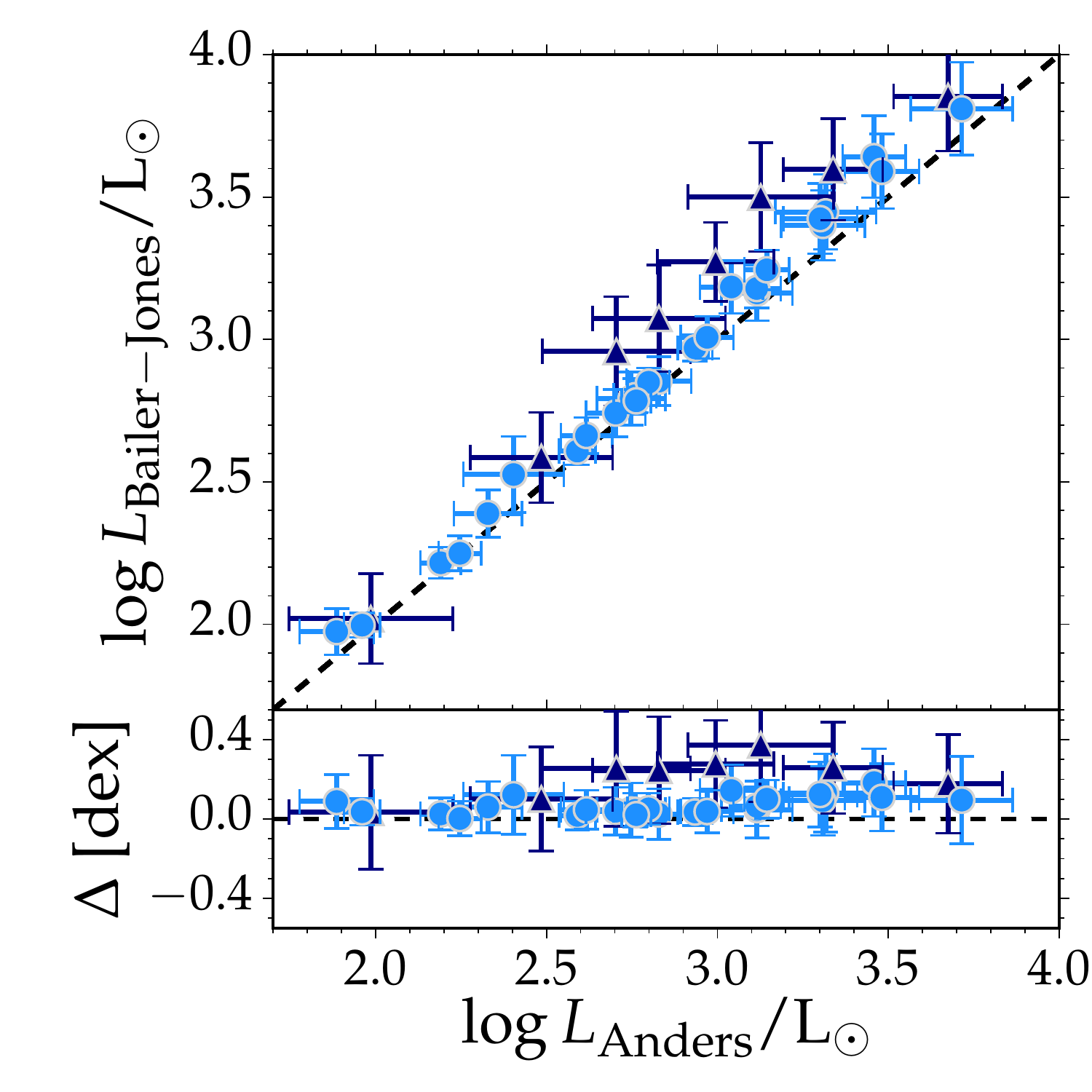}
	\caption{\textcolor{black}{Comparison} of luminsoties derived using the distances from \citet[][x-axis]{Anders2019} and \citet[][y-axis]{Bailer-Jones2018}. Dark triangles mark the stars for which the error of the parallaxes are $\geq 20$ per cent. Residuals are shown in the bottom panel. \textcolor{black}{The plotted data on the y-axis are listed in the third column of Table~G2.}}
	\label{fig:Lum_comp2}
\end{figure}

The differences in luminosities arising from using the distances from \citet[][$L_\text{Bailer-Jones}$]{Bailer-Jones2018} instead of those from \citet[][$L_\text{Anders}$]{Anders2019} are illustrated in the left panel of Fig.\,\ref{fig:Lum_comp2}. Stars for which the errors on the parallaxes are  $\geq 20$ per cent are indicated by dark triangles. The \citet{Bailer-Jones2018} luminosities are generally found to be larger than the $L_\text{Anders}$ values, with the discrepancies tending to be larger for higher luminosities and for stars with $\varpi_\text{error} > 20$ per cent. For such high errors on the parallaxes, the derived distances are mainly defined by the selected priors in the derivations of the distances \citep{Bailer-Jones2018}. \textcolor{black}{The choice of priors is} the main source behind the discrepancies between \citet{Bailer-Jones2018} and \citet{Anders2019} distances and thereby the derived luminosities. \textcolor{black}{F}or all 34~SPB stars the distances from \citet{Bailer-Jones2018} are larger than those from \citet{Anders2019}, \textcolor{black}{contrary to the general findings by \citet{Anders2019}.} In the end, we find that the differences between $L_\text{Bailer-Jones}$ and $L_\text{Anders}$ to be in the range of \textcolor{black}{0.00}--0.37\,dex.

%%%%%%%%%%%%%%%%%%%%%%%%%%%%%%%%%%%%%%%%%%%%%%%%%%%%%%%%%%%%%%%%%%%%%%%%%%%%%%%%%%%
%				Placement in the SPB instability strip
%%%%%%%%%%%%%%%%%%%%%%%%%%%%%%%%%%%%%%%%%%%%%%%%%%%%%%%%%%%%%%%%%%%%%%%%%%%%%%%%%%%
%\clearpage
\section{The SPB instability strip}\label{Sec:HRD}

\begin{figure*}%[h]
  \centering
  \includegraphics[width=0.75\linewidth]{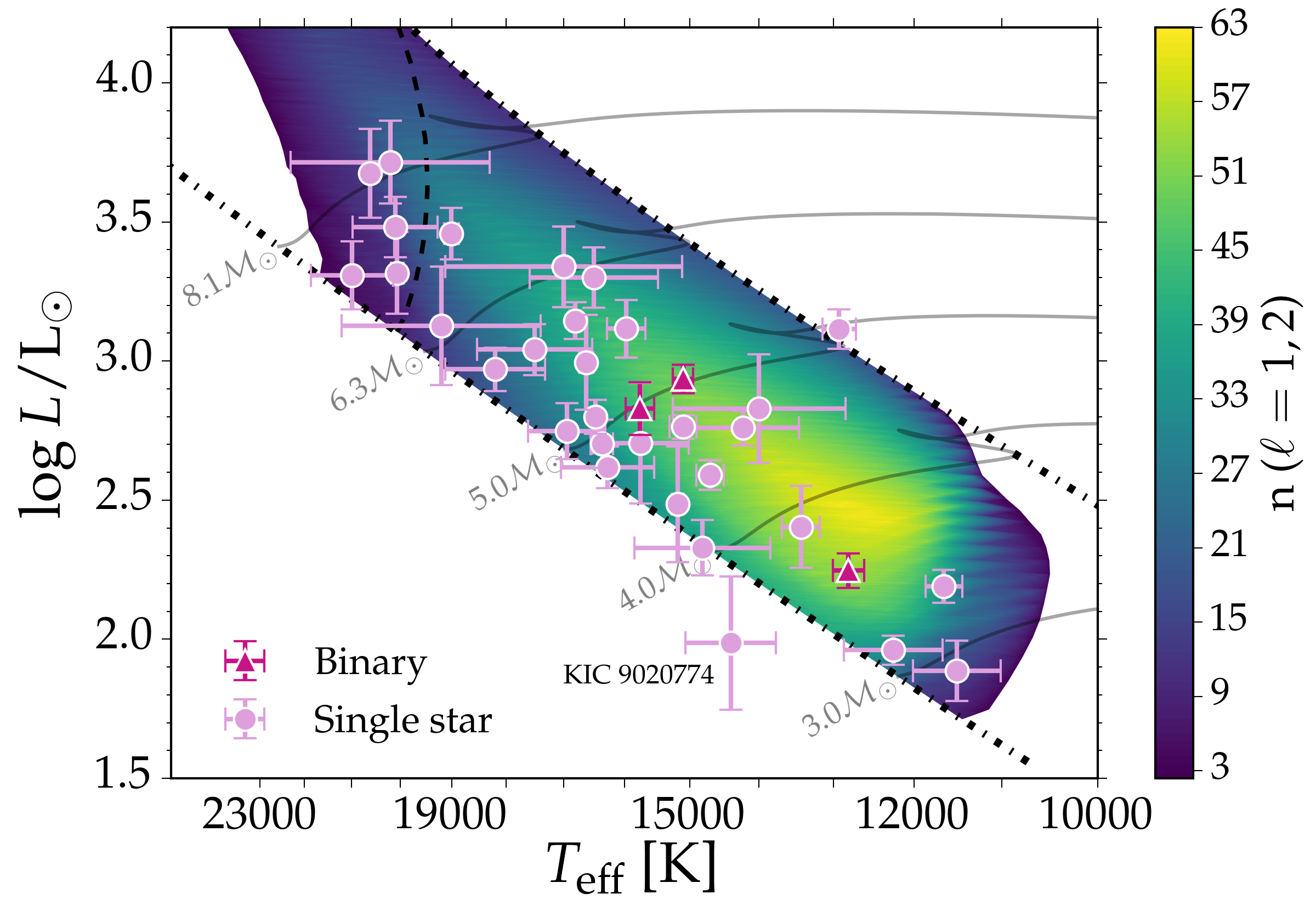}
\caption{\textcolor{black}{HR} diagram showing \textcolor{black}{the 34~SPB} stars in the SPB instability strip. \textcolor{black}{Binaries} are marked by dark triangles. The instability strip \textcolor{black}{by \citet{Moravveji2016} assumes} a metallicity of $Z = 0.014$, the metal mixture of \citet{Asplund2009}, an exponential convective core overshooting of $f_\text{ov} = 0.02$, and an 75\% increase in the nickle and iron opacities. The coloured region shows the sum of the number \textcolor{black}{$n$} of excited dipole ($\ell = 1$) and quadruple ($\ell = 2$) gravity modes\textcolor{black}{.} The top and bottom dot-dashed lines indicates the position of the zero-age and terminal-age main-sequence, while the black dashed line shows the cool edge of the $\beta$ \textcolor{black}{Cep} instability strip for radial ($\ell = 0$) pressure modes. Five example evolutionary tracks are shown in grey and labeled according to their initial mass.}
	\label{fig:HRD}
\end{figure*}

Using the luminosities labeled as $L_\star$ in Table\,G1, we place the 34~SPB stars in the \textcolor{black}{HR} diagram and compare their positions to the predicted SPB instability strip derived by \citet{Moravveji2016}, see Fig.\,\ref{fig:HRD}. The colours of the instability strip \textcolor{black}{indicate} the expected \textcolor{black}{number} of excited dipole and quadruple gravity modes, with brighter regions corresponding to a higher number of \textcolor{black}{excited modes. Most} of the stars appear to have masses higher than $4\ \text{M}_\odot$. Five of the stars also fall within the cool edge of the $\beta$ \textcolor{black}{Cep} instability strip indicated by the black dashed line in Fig.\,\ref{fig:HRD}, and are therefore predicted to be hybrid pulsators showing \textcolor{black}{both gravity} and pressure modes. 

\begin{figure}%[h]
  \centering
  \includegraphics[width=\linewidth]{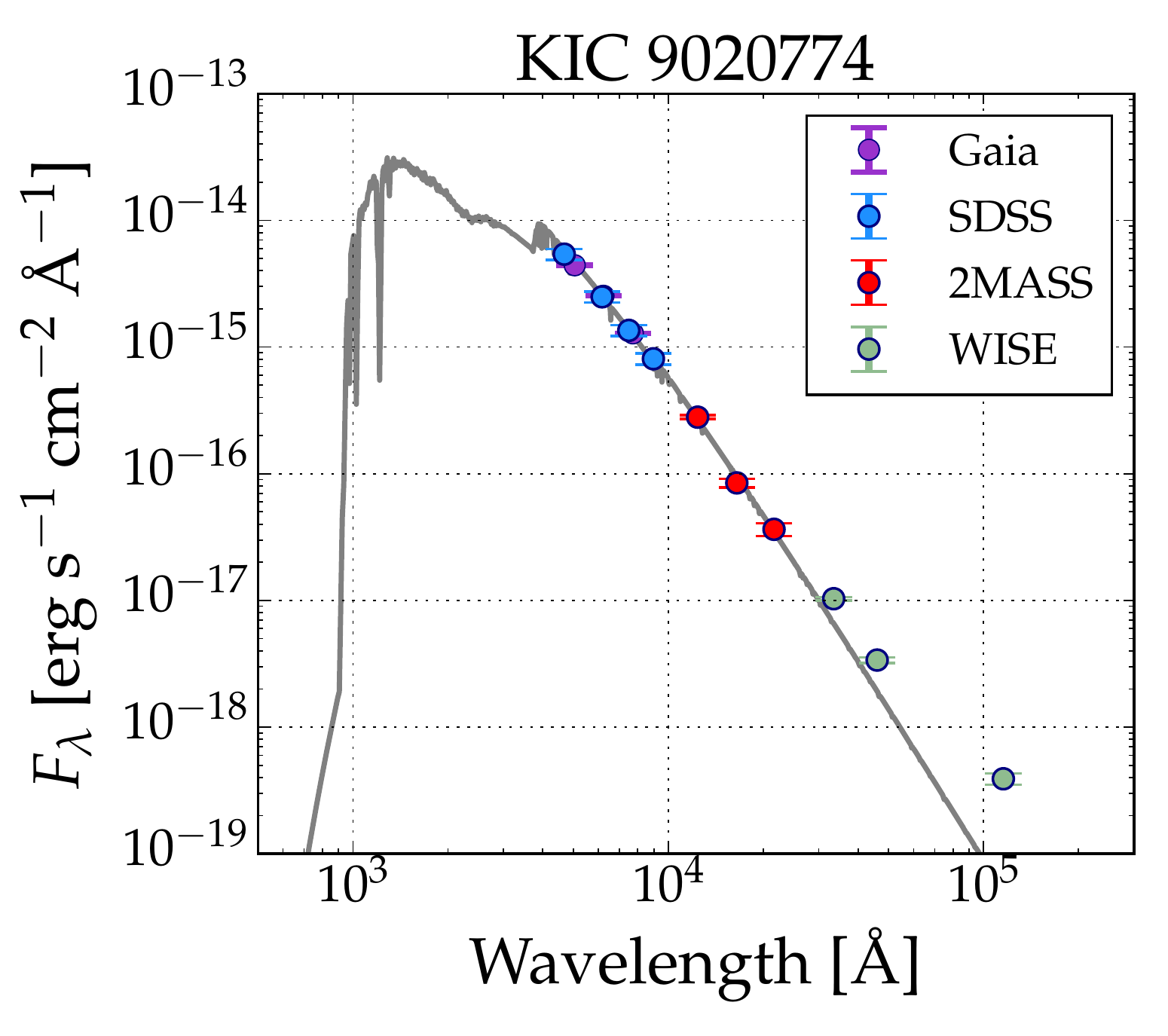}
\caption{Fitted spectral energy distribution of KIC~9020774 compared to the measured photometry in the \emph{Gaia}, SDSS, 2MASS and WISE passbands. An infrared excess is seen in the WISE photometry.}
	\label{fig:KIC90_SED}
\end{figure}

\textcolor{black}{One outlier} shows up in the HR diagram. The derived luminosity of KIC~9020774 is too low to place it inside the SPB instability, and also causes it to fall below the main-sequence. \textcolor{black}{When} plotted in the Kiel diagram\footnote{Surface gravity vs effective temperature.} this is no longer the case as seen in Fig.\,1 from \citet{Papics2017}. To investigate the possible cause of this discrepancy, we plot its fitted spectral energy distribution against its measured flux densities $F_\lambda$ in the \emph{Gaia}, SDSS, 2MASS, and WISE filters in Fig.\,\ref{fig:KIC90_SED}. The SED is obtained by fixing the $T_\text{eff}$, $\log g$, [M/H], and $E(B-V)$ of the star to its spectroscopic values and the results from the \textsf{Bayestar19} reddening map, and afterwards varying the angular diameter $\alpha$ until the best match to the observations is obtained, following the procedures outlined by \citet{Degroote2011}. The final fitted SED is shown in grey in Fig.\,\ref{fig:KIC90_SED}.

Comparing the broadband photometric data to the theoretical SED, an infrared excess is observed for all of the WISE measurements. As previously mention\textcolor{black}{ed} in Sect.\,\ref{Sec:Lum_from_BC}, such an excess can be expected when circumstellar dust is present. \textcolor{black}{This} may cause the total reddening of the star to be underestimated, and lead to further underestimation of the actual luminosity of the star. We estimate that a circumstellar reddening of $E(B-V)_\text{C} \sim 0.23$ would be needed to move KIC~9020774 fully within the SPB instability \textcolor{black}{strip.} Keeping in mind the position of KIC~9020774 in the Kiel diagram, we hypothesize that the star might be a very young main-sequence star which still has circumstellar material left from its pre-main-sequence phase.

%%%%%%%%%%%%%%%%%%%%%%%%%%%%%%%%%%%%%%%%%%%%%%%%%%%%%%%%%%%%%%%%%%%%%%%%%%%%%%%%%%%
%				Conclusions
%%%%%%%%%%%%%%%%%%%%%%%%%%%%%%%%%%%%%%%%%%%%%%%%%%%%%%%%%%%%%%%%%%%%%%%%%%%%%%%%%%%

\section{Conclusions}\label{Sec:Conclusions}

Starting from two grids of model atmospheres assuming LTE and NLTE, respectively, we have calculated an LTE and combined LTE+NLTE grid of bolometric corrections. Through the use of multivariate linear regression, three statistical model representations of the LTE and LTE+NLTE grids were computed for each of the considered 27 filters as a function of $T_\text{eff}$, $\log g$, and [M/H]. These prescriptions are valid for $T_\text{eff} \in [10000, 30000]$\,K and \textcolor{black}{for} a wider range in $\log g$ and [M/H] than a similar prescription for BC$_V$ by, e.g., \citet{Flower1996}, which has been derived for main-sequence stars. \textcolor{black}{A high microturbulence ($V_t = 10$\,km\,s$^{-1}$) combined with low surface gravity ($\log g \leq 3.0$) may result in an underestimation of the bolometric corrections derived from the three prescriptions presented in this work.}

Using these derived prescriptions for the bolometric corrections for each passband and the \textcolor{black}{\textsf{Bayestar19} reddening map by \citet{Green2019}}, we calculate an average, extinction corrected, apparent bolometric magnitude for each of the 34~SPB stars. These are converted to luminosities using the distances from \citet{Bailer-Jones2018} and \citet{Anders2019} based on the \emph{Gaia} DR2 parallaxes. We find that excluding NLTE effects has no significant impact on the luminosities within the temperature range of the considered SPB stars, and that the same luminosities \textcolor{black}{are obtained} if the bolometric corrections are derived using a grid interpolation. If the BC$_V$ vs $\log T_\text{eff}$ prescription by \citet{Flower1996} is used, the luminosities remain within the errors of those derived from our statistical models for the majority of the stars. The largest discrepancies in the luminosities is obtained when the stellar parameters from \emph{Gaia} \citep{Anders2019} or the \emph{Kepler Input Catalog} are used when calculating the bolometric corrections. In these cases, differences as large as \textcolor{black}{0.59} and \textcolor{black}{0.66}\,dex in luminosity are reached for the \emph{Gaia} and KIC parameters, respectively. Furthermore, \textcolor{black}{the} luminosities are larger when the \citet{Bailer-Jones2018} distances are used instead of the ones from \citet{Anders2019}\textcolor{black}{.} 

The derived luminosities were used to place the 34~SPB stars in the HR diagram and compare their positions with the theoretical SPB instability computed by \citet{Moravveji2016}. \textcolor{black}{While} the three statistical model prescriptions for the bolometric corrections in difference passbands were derived in order to obtain accurate luminosities of the 34~SPB stars, \textcolor{black}{this} is only one out of many possible applications for B-type stars. \textcolor{black}{Our statistical recipes to compute the bolometric corrections for the numerous photometric filters considered in this work are readily available for other astrophysical applications.}

\section*{Acknowledgements}

The research leading to these results has received funding from the European Research Council (ERC) under the European Union's Horizon 2020 research and innovation programme (grant agreement N$^\circ$670519: MAMSIE) and from the KU\,Leuven Research Council (grant C16/18/005: PARADISE). AE acknowledges support from the Fonds voor Wetenschappelijk Onderzoek Vlaanderen (FWO) under contract ZKD1501-00-W01.
Furthermore, this work has made use of data from the European Space Agency (ESA) mission Gaia (https://www.cosmos.esa.int/gaia), processed by the Gaia Data Processing and Analysis Consortium (DPAC, https://www.cosmos.esa.int/web/gaia/dpac/consortium). Funding for the DPAC has been provided by national institutions, in particular the institutions participating in the Gaia Multilateral Agreement. 
This publication makes use of data products from the Two Micron All Sky Survey, which is a joint project of the University of Massachusetts and the Infrared Processing and Analysis Center/California Institute of Technology, funded by the National Aeronautics and Space Administration and the National Science Foundation. 

%%%%%%%%%%%%%%%%%%%%%%%%%%%%%%%%%%%%%%%%%%%%%%%%%%



\bibliographystyle{mnras}
\bibliography{references.bib} % if your bibtex file is called example.bib

%%%%%%%%%%%%%%%%%%%%%%%%%%%%%%%%%%%%%%%%%%%%%%%%%%
% Don't change these lines
\bsp	% typesetting comment
\label{lastpage}
\end{document}